\documentclass[11pt]{article}
\usepackage{amsmath}
\usepackage{amsthm}
\usepackage{amsfonts}
\usepackage[english]{babel}
\usepackage{amssymb}
\usepackage{graphicx}
\usepackage[a4paper]{geometry}
\usepackage{hyperref}
\usepackage{color}

\newcommand{\be}{\begin{equation}}
\newcommand{\ee}{\end{equation}}
\newcommand{\bea}{\begin{eqnarray}}
\newcommand{\eea}{\end{eqnarray}}
\newcommand{\vsp}{\vspace{0.4cm}}

\newcommand{\appa}{\mathcal{A}}
\newcommand{\stsp}{\mathcal{S}}
\newcommand{\obsp}{\mathfrak{O}}

\newtheorem{rem}{Remark}

\newtheorem{defn}{Definition}

\newtheorem{exmp}{Example}

\newtheorem{thm}{Theorem}

\newtheorem{prop}{Proposition}

\newtheorem*{proposition*}{Proposition}

\newtheorem*{pf}{Proof}

\title{Dynamical Vector Fields on the Manifold of Quantum States}
\author{F. M. Ciaglia$^{1}$ $^{2}$, F. Di Cosmo$^{1}$ $^{2}$, A. Ibort$^{3}$ $^{4}$, M. Laudato $^{1}$, G. Marmo$^{1}$ $^{2}$\\
\textit{$^{1}$ Dipartimento di Fisica, Universit\`a di Napoli ``Federico II",}\\
\textit{Via Cinthia Edificio 6, I-80126 Napoli, Italy}\\
\textit{$^{2}$ INFN-Sezione di Napoli, Via Cinthia Edificio 6, I-80126 Napoli, Italy}\\
\textit{$^{3}$ Departamento de Matem\'{a}ticas, Universidad Carlos III de Madrid}\\
\textit{Avda. de la Universidad 30, 28911 Legan\'{e}s, Madrid, Spain.}\\
\textit{$^{4}$ ICMAT, Instituto de Ciencias Matem\'{a}ticas (CSIC - UAM - UC3M - UCM)} \\  
\textit{Nicol\'{a}s Cabrera,13–15, Campus de Cantoblanco, UAM, 28049, Madrid, Spain}}
\date{}

\begin{document}

\maketitle

\abstract{In this paper we shall consider the stratified manifold of quantum states and the vector fields which act on it.
In particular, we show that the infinitesimal generator of the GKLS evolution is composed of a generator of unitary transformations plus a gradient vector field along with a Kraus vector field transversal to the strata defined by the involutive distribution generated by the former ones.}

\section{Introduction}

The mathematical description of (Markovian) open quantum systems was initiated in the pioneriing works \cite{gorini_kossakowski_sudarshan-completely_positive_dynamical_semigroups_of_N-level_systems} and \cite{lindblad-on_the_generators_of_quantum_dynamical_semigroups}.
In these papers, the explicit form of the most general master equation governing the Markovian dynamics of a finite-level quantum system was found.

Despite the ``absolute'' character of this result, the theoretical and experimental richness of the theory of open quantum systems is continuously growing.
The important increase in the level of experimental control on quantum systems has led to a wide number of experimental realizations of open quantum systems in different fields of physical applications.
For instance, in quantum optics; in atomic and molecular physics; and in mesoscopic physics.

\vsp

An open system can be thought of as a physical system $\mathit{S}$ which is not closed, that is, it is interacting in some way with an environment $\mathit{E}$.
From the conceptual point of view, one may hope to be able to consider a new physical system $\mathit{T}$, which is the sum of $\mathit{S}$ and $\mathit{E}$, so that $\mathit{T}$ becomes a closed system.

This conceptual attitude is corroborated by a number of mathematical results, both in classical, and quantum physics.
Indeed, given a classical system $\mathit{S}$ the dynamical evolution of which is described by means of a vector field $X$ on some carrier manifold $M$, it is always possible to find a symplectic lift $\tilde{X}$ of $X$ to the cotangent bundle $T^{*}M$, so that $\tilde{X}$ becomes a Hamiltonian vector field, that is, $\tilde{X}$ describes a closed classical system \cite{arnold-mathematical_methods_of_classical_mechanics,abraham_marsden-foundations_of_mechanics}.

On the other hand, the state $\rho$ of a (finite-level) quantum system $\mathit{S}$, is described by a density matrix in $\mathcal{B}(\mathcal{H}_{\mathit{S}})$, where $\mathcal{H}_{\mathit{S}}$ is the Hilbert space of the system.
Here, the evolution of a closed system corresponds to the unitary evolution:

\be
\Phi_{\tau}(\rho)=\mathbf{U}_{\tau}\,\rho\,\mathbf{U}_{\tau}^{\dagger}\,,
\ee
with $\mathbf{U}_{\tau}$ a unitary operator for all $\tau$.
An open system is described by a semigroup $\Phi_{\tau}$ of completely-positive trace preserving (CPTP) maps from $\mathcal{B}(\mathcal{H}_{\mathit{S}})$ into itself.
In this context, Stinespring theorem \cite{stinespring-positive_functions_on_cstar_algebras} states that every completely-positive trace-preserving map $\mathcal{K}(\rho)$ from $\mathcal{B}(\mathcal{H}_{\mathit{S}})$ to itself can be obtained in three steps.
First, we have to consider the tensor product $\mathcal{H}_{\mathit{S}}$ with an auxiliary Hilbert space $\mathcal{H}_{\mathit{E}}$.
According to the general postulates of quantum mechanics \cite{dirac-principles_of_quantum_mechanics}, $\mathcal{H}_{\mathit{S}}\otimes\mathcal{H}_{\mathit{E}}$ represents the Hilbert space of the composite system $\mathit{S} + \mathit{E}$.
Once we have the composite system, we let it evolve  by means of a unitary evolution depending on the explicit form of $\mathcal{K}$.
Finally, we project the evolved state back from $\mathcal{H}_{\mathit{S}}\otimes\mathcal{H}_{\mathit{E}}$ to $\mathcal{H}_{\mathit{S}}$ to define an evolution from $\mathcal{H}_{\mathit{S}}$ into $\mathcal{H}_{\mathit{S}}$.

\vsp

Although these prescription seems clear cut, its practical implementation suffers of some limitations.
Indeed, it is often the case that the environment is so complicated that a complete knowledge of the actual state describing it is impossible.
Consequently, it is impossible to determine the evolution of the composite system because we do not know the initial state of the composite system.
Furthermore, it is very likely that our knowledge of the explicit form of the interaction between the system and its environment is   uknown to us.
What we actually have, is only an effective dynamics on the subsystem $\mathit{S}$.
The Gorini-Kossakowski-Sudarshan-Linbland equation (or GKLS for short) \cite{gorini_kossakowski_sudarshan-completely_positive_dynamical_semigroups_of_N-level_systems,lindblad-on_the_generators_of_quantum_dynamical_semigroups}:

\be\label{eqn: GKLS equation introduction}
\mathbf{L}(\rho)=-\imath\left[\mathbf{H}\,,\rho\right] - \frac{1}{2}\sum_{j=1}^{N}\,\left\{\mathbf{v}_{j}^{\dagger}\mathbf{v}_{j}\,,\rho\right\} + \sum_{j=1}^{N}\,\mathbf{v}_{j}\,\rho\,\mathbf{v}_{j}^{\dagger}\,
\ee
describes precisely the most general form for the generator of a finite-level open quantum system from the perspective of the effective dynamics.

\vsp

From the mathematical point of view, equation (\ref{eqn: GKLS equation introduction}) has a clear algebraic flavour.
This follows from the fact that the most used mathematical tools in quantum mechanics are algebraic.
However, in the last decades, something changed, and the geometrical picture of quantum mechanics has started to grow \cite{ashtekar_schilling-geometrical_formulation_of_quantum_mechanics,cirelli_mania_pizzocchero-quantum_mechanics_as_an_infinite_dimensional_Hamiltonian_system_with_uncertainty_structure,ercolessi_marmo_morandi-from_the_equations_of_motion_to_the_canonical_commutation_relations,carinena_clemente-gallardo_marmo-geometrization_of_quantum_mechanics,chruscinski_jamiolkowski-geometric_phases_in_classical_and_quantum_mechanics,bengtsson_zyczkowski-geometry_of_quantum_states:_an_introduction_to_quantum_entanglement}.

In this picture, a rich geometrical structure associated with finite-level quantum systems naturally emerges.
For instance, denoting with $\sigma$ the spectrum of the density matrix associated with a quantum state $\rho$, that is, the eigenvalues of $\rho$, the set $\stsp_{\sigma}$ of all  quantum states with the same spectrum $\sigma$ turns out to be a K\"{a}hler manifold.
In particular, the space of pure states is the complex projective space\footnote{The set $P(\mathcal{H})$ is a K\"{a}hler manifold even in the infinite-dimensional case \cite{cirelli_lanzavecchia_mania-normal_pure_states_and_the_von_neumann_algebra_of_bounded_operators_as_kahler_manifold}.} $P(\mathcal{H})$, which is a well-known K\"{a}hler manifold.
More generally, for all $k=1\,,...,n$ where $n=dim(\mathcal{H})$, the set $\stsp_{k}$ of all quantum states with rank equal to $k$, is a homogeneous space for the natural action of the special linear group $SL(\mathcal{H}\,,\mathbb{C})$ \cite{grabowski_kus_marmo-symmetries_group_actions_and_entanglement,grabowski_kus_marmo-geometry_of_quantum_systems_density_states_and_entanglement}, and thus, a differential manifold.

Unitary evolutions are realized by means of Hamiltonian vector fields on the manifold $\stsp_{\sigma}$ of isospectral states.
However, open quantum dynamics may change both the spectrum and the rank of a quantum state, and thus the geometrical description of such dynamical processes can not be accomplished resorting to  the differential structure of $\stsp_{\sigma}$ or $\stsp_{k}$.
We must be able to describe the motion across orbits of quantum states of different rank, and this is precisely the aim of this paper.

\vsp

We will give a geometrical formulation of the dynamics of open quantum systems generated by the GKLS operator $\mathbf{L}$ of equation (\ref{eqn: GKLS equation introduction}) in the case of finite-level quantum systems.
Specifically, we will describe the dynamics of open quantum systems determined by GKSL equation by means of a vector field $\Gamma$ on a suitable differential manifold.
At this purpose, we note that the dynamical trajectories of quantum states under open quantum dynamics lies entirely in the set $\mathfrak{T}_{1}$ of self-adjoint operators with trace equal to 1.
Consequently, we will rely on the differential structure of $\mathfrak{T}_{1}$ in order to describe open quantum dynamics by means of vector field $\Gamma$ which will turn out to be a fine-tuned combination of geometrically meaningful vector fields.
Specifically, we will get a decomposition of $\Gamma$ as (compare with the three terms in the r.h.s. of equation (\ref{eqn: GKLS equation introduction})):

\be
\Gamma= X + Y + Z\,,
\ee
where $X$ is a Hamiltonian vector field the flow of which preserves the spectrum of quantum states, $Y$ is a gradient-like vector field whose flow changes the spectrum but preserves the rank of quantum states, and $Z$ is a vector field the flow of which is responsible for the change in rank of quantum states.
Interestingly, $X$ will turn out to be an affine vector field which need not be correlated with $Y$ and $Z$.
On the other hand, $Y$ and $Z$ will turn out to be highly related.
They will be non-affine vector fields such that their sum is an affine vector field.

To accomplish this task, we will make use of the Lie-Jordan algebra structure on the space of linear functions on the dual $\obsp^{*}$ of the space of self-adjoint operators $\obsp$, and exploit  a symmetric and an anti-symmetric product structure on the algebra $\mathcal{F}(\mathfrak{T}_{1})$ of smooth functions on $\mathfrak{T}_{1}$.
These products allow us to define, respectively,  the gradient-like and the Hamiltonian vector field by means of the affine functions on $\mathcal{F}(\mathfrak{T}_{1})$ associated with elements of $\obsp$.
By construction, these vector fields will be precisely the vector fields generating the nonlinear action of $SL(\mathcal{H}\,,\mathbb{C})$ of which all spaces $\stsp_{k}$ are homogeneous spaces.
Consequently, the trajectory of a quantum state $\rho\in\stsp_{k}$ by means of the flow of these vector fields will be completely contained in $\stsp_{k}$.
The vector field $Z$ will be constructed with the help of an affine map on $\mathfrak{T}_{1}$.

\vsp

Some similar ideas are exposed in \cite{carinena_clemente-gallardo_jover-galtier_marmo-tensorial_dynamics_on_the_space_of_quantum_states,ciaglia_dicosmo_laudato_marmo-differential_calculus_on_manifolds_with_boundary.applications} using a different mathematical perspective.
Indeed, the ambient space used in \cite{carinena_clemente-gallardo_jover-galtier_marmo-tensorial_dynamics_on_the_space_of_quantum_states,ciaglia_dicosmo_laudato_marmo-differential_calculus_on_manifolds_with_boundary.applications} is not $\mathfrak{T}_{1}$, but rather the space $\obsp^{*}$ of positive functionals on the $C^{*}$-algebra $\appa=\mathcal{B}(\mathcal{H})$ of the system.
In this picture, the normalization of a quantum state is taken into account imposing an ad-hoc constraint on the vector field representing the dynamics.
A similar ad-hoc constraint is imposed in order to define the bivector fields by means of which the Hamiltonian and gradient-like vector fields are constructed.
On the other hand, the formalism presented here does not need any ad-hoc constraint because the ambient space $\mathfrak{T}_{1}$ already takes into account the normalization of a quantum state.
Furthermore, the bivector fields giving rise to the Hamiltonian and gradient-like vector fields are introduced here using an abstract procedure of reduction for product structures on algebras of smooth functions.

\vsp

Once we have this geometrical formulation of open quantum dynamics, some interesting possibile applications arise.
Indeed, the mathematical results of the theory of dynamical systems, which are mainly related to the geometrical structure of classical mechanics,  become immediately available in the quantum case because of the common mathematical language in which classical physics and open quantum dynamics are here formulated, namely, using vector fields on differential manifolds.
  
We believe that the interplay between the mathematical methods of classical physics and the quantum theory could help to  better understand the structure underlying quantum physics, and to provide some useful tools in the computation of specific physical situations.

Of course, we are not saying that classical physics should drive our understanding of quantum physics.
We are simply pointing out how casting physical problems pertaining to the quantum domain into a mathematical formalism which is common to classical physics leads us to benefit of all the mathematical results available in that formalism.
Clearly, the physical interpretation of these results must be consistent with the quantum nature of the system at hand.
A similar attitude, but in the opposite direction, was pursued by Koopman \cite{koopman-hamiltonian_systems_and_transformations_in_hilbert_space} who reformulated the dynamical problem of classical physics in the mathematical formalism of Hilbert spaces characteristic of quantum mechanics.

\vsp

The article is organized as follows.
In section \ref{sec: Geometry of quantum states}, we review the geometrical structure of finite-level quantum systems and introduce  the mathematical tools we need to construct the GKLS vector field $\Gamma$.
In section \ref{sec:GKLS vector field}, we actually show how to construct the GKLS vector field $\Gamma$ representing the geometrical version of the GKLS generator $\mathbf{L}$ in equation (\ref{eqn: GKLS equation introduction}).
In section \ref{sec: Quantum Poisson semigroups}, the geometrical formulation of open quantum dynamics is applied to the some class of dynamics, specifically, the so-called quantum Poisson semigroups, the so-called quantum Gaussian semigroups,
 and the so-called random unitary semigroups \cite{lindblad-on_the_generators_of_quantum_dynamical_semigroups,kossakowski-on_quantum_statistical_mechanics_of_non_hamiltonian_systems,aniello_kossakowski_marmo_ventriglia-quantum_brownian_motion_on_lie_groups_and_open_quantum_systems}.
By using a variation of LaSalle principle, what is found is that, in every dimension, all these dynamical systems present an attractor in the sense of dynamical systems, i.e., the dynamical evolution of every quantum state $\rho$ tends to a non-equilibrium steady state $\rho_{\infty}$ contained in a set $S_{\infty}$ which is independent of the initial state $\rho$.
The set $S_{\infty}$ is the largest invariant subset in the intersection of the space $\stsp$ of quantum states with the set:

\be
E:=\left\{\rho\in\stsp\colon \left.\mathcal{L}_{\Gamma}\chi\right|_{\rho}=0\right\}\,,
\ee
where $\chi$ is the purity function, and $L_{\Gamma}\chi$ denotes the Lie derivative of $\chi$ with respect to the GKLS vector field $\Gamma$.

\section{Geometry of quantum states}\label{sec: Geometry of quantum states}

Let $\appa=\mathcal{B}(\mathcal{H})$ denote the $C^{*}$-algebra of  all bounded operators on the Hilbert space $\mathcal{H}$ of a quantum system.  In what follows we will assume for simplicity that we are dealing with a finite dimensional Hilbert space $\mathcal{H}\cong\mathbb{C}^{n}$.
Let $\obsp$ denote the space of self-adjoint elements in $\appa$.

The space $\stsp$ of quantum states of the system, i.e., positive normalized linear functionals on $\mathcal{A}$, is a compact convex body in $\obsp^{*}$ \cite{bratteli_robinson-operator_algebras_and_quantum_statistical_mechanics_1,dixmier-c*_algebras}.
Except for $n=2$, $\stsp$ lacks of a differential structure as a whole because its boundary is not a smooth submanifold.
However, it is a subset of the affine space $\mathfrak{T}_{1}$ of elements  $\omega\in\obsp^{*}$ such that $\omega(\mathbb{I})=1$, where $\mathbb{I}\in\appa$ is the identity operator.
This is an affine subspace of $\obsp^{*}$ carrying the structure of an embedded submanifold, being it the inverse image of the closed set $\{1\}$ by means of the linear function $f_{\mathbb{I}}(\omega)=\omega(\mathbb{I})$.

Being  $\mathcal{H}\cong\mathbb{C}^{n}$, we may realize $\appa$ as the space of $(n\times n)$ complex matrices.
Accordingly, elements in $\obsp$ are realized as $(n\times n)$ self-adjoint matrices.
 
In the finite-dimensional case, $\obsp$ is isomorphic to its dual, and thus, every $\omega\in\obsp^{*}$ may  be realized as a $(n\times n)$  self-adjoint matrix.
The duality between $\mathbf{a}\in\obsp$ and $\omega\in\obsp^{*}$ is expressed using the matrix trace, specifically:

\be
\omega(\mathbf{a})=Tr(\omega\,\mathbf{a})\,,
\ee
where, by an abuse of notation, $\omega$ and $\mathbf{a}$ in the right hand side denote the matrix expressions of $\omega$ and $\mathbf{a}$ in the left hand side.
In this matrix picture, the space $\stsp$ of quantum states consists of $(n\times n)$ positive matrices with trace equal to $1$.

In the following, we will  often use these matrix representations as an effective way to conveniently perform calculations using matrix algebra.

We may define a nonlinear action $\alpha$ of the special linear group $SL(\mathcal{H}\,,\mathbb{C})$ on the positive elements of $\mathfrak{T}_{1}$ setting:

\be\label{eq: nonlinear action of SL(n,C)}
(\mathbf{g}\,,\rho)\mapsto \alpha_{\mathbf{g}}(\rho):=\frac{\mathbf{g}\,\rho\,\mathbf{g}^{\dagger}}{Tr(\mathbf{g}\,\rho\,\mathbf{g}^{\dagger})}\,.
\ee
Note that this action is well defined only on positive or negative elements of $\mathfrak{T}_{1}$.
Indeed, when $\xi\in\mathfrak{T}_{1}$ has positive and negative eigenvalues, then there is $\mathbf{g}$ such that $Tr(\mathbf{g}\,\xi\,\mathbf{g}^{\dagger})=0$, and thus, the denominator in equation (\ref{eq: nonlinear action of SL(n,C)}) blows up.

\begin{rem}
Let us look at consider a concrete example of a two-dimensional quantum system.
An element $\xi\in\obsp^{*}$ given by:

\be
\xi=x^{0}\,\sigma_{0} + x^{1}\sigma_{1}\,,
\ee
where the $\{\sigma_{\mu}\}$ are the Pauli matrices with $\sigma_{0}=\mathbb{I}$ the identity operator.
Note that the Pauli matrices are a basis for the Lie algebra of $GL(2\,,\mathbb{C})$.

\begin{figure}[h!]
\centering
\includegraphics[scale=0.3]{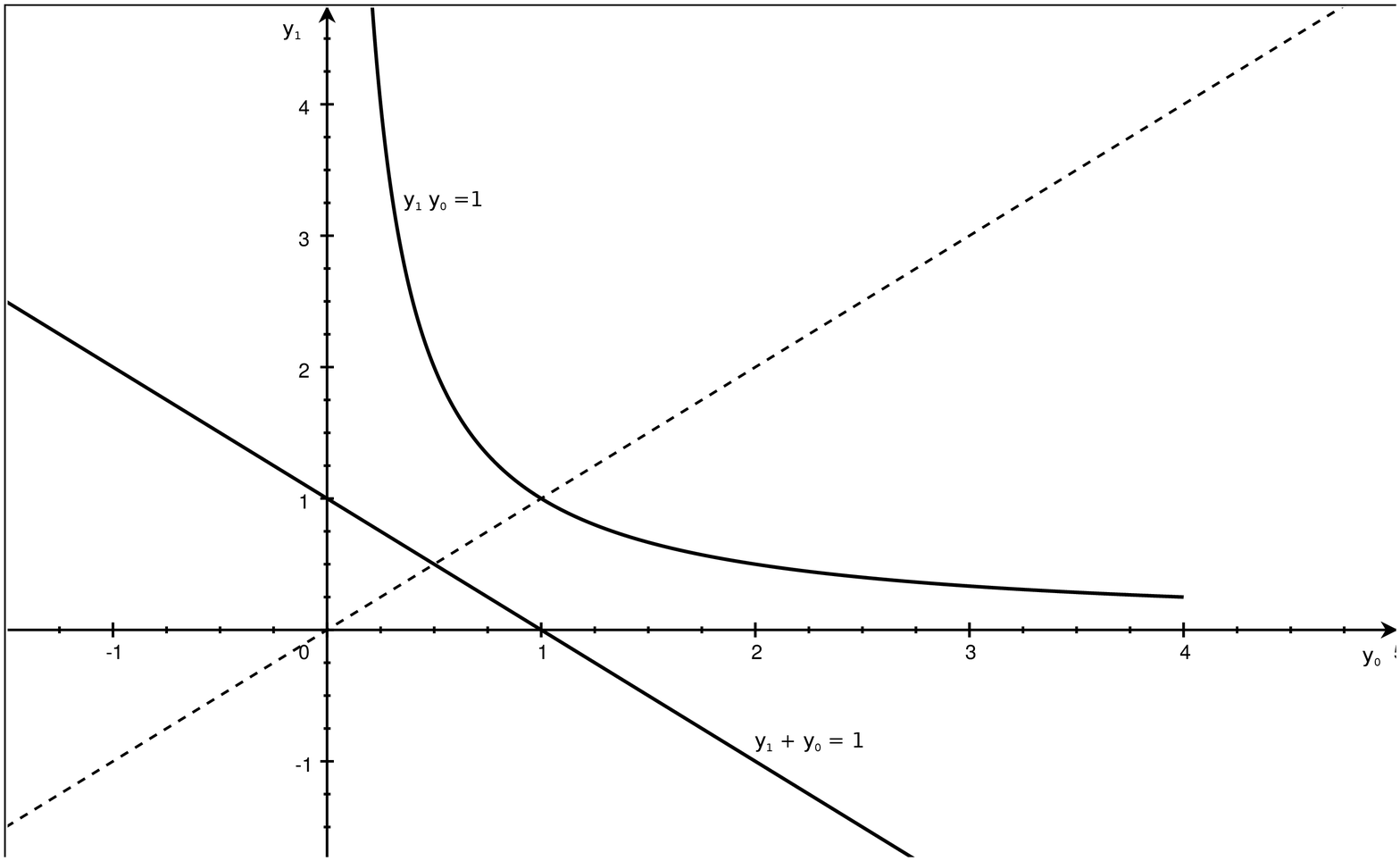}
\caption{\textit{}}
\label{Figure1}
\end{figure}

Let $\mathbf{g}\in SL(2\,,\mathbb{C})$ be of the form

\be
\mathbf{g}=\mathrm{e}^{\mathbf{A}}=\mathrm{e}^{a\sigma_{1}}=\cosh(a)\mathbb{I} + \sinh(a)\sigma_{1}\,.
\ee
It is clear that $det(\xi)=\det(\mathbf{g}\,\xi\,\mathbf{g}^{\dagger})$.
Furthermore, it is $det(\xi)=(x^{0})^{2} - (x^{1})^{2}$.
Let us now perform the following change of coordinates $y^{0}=x^{0} + x^{1}$, $y^{1}=x^{0} - x^{1}$, so that $\det(\xi)=y^{0}\,y^{1}$.
Accordingly, in the $(y^{0}\,,y^{1})$-plane,  every  element $\xi$ with $\det(\xi)=c$ is represented as a point on the hyperboloid $y^{0}\,y^{1}=c$, and the action $\xi\mapsto \mathbf{g}\,\xi\,\mathbf{g}^{\dagger}$ moves the point on this hyperboloid.

\begin{figure}[h!]
\centering
\includegraphics[scale=0.3]{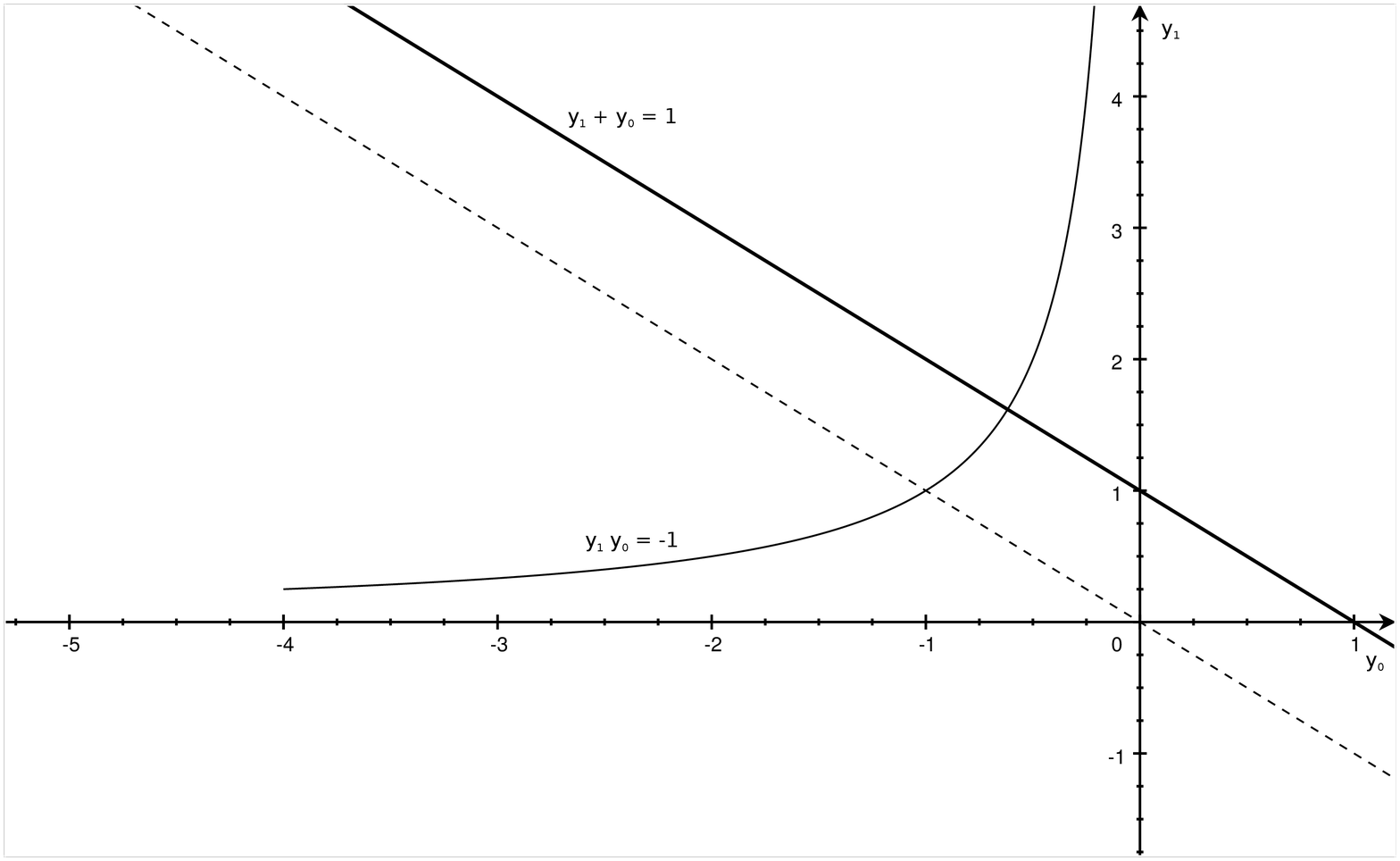}
\caption{\textit{}}
\label{Figure2}
\end{figure}

In this framework, to divide $\xi_{g}=\mathbf{g}\,\xi\,\mathbf{g}^{\dagger}$ by its trace $Tr(\mathbf{g}\,\xi\,\mathbf{g}^{\dagger})$ corresponds to move the point $\xi_{g}$ on the hyperboloid to the point $\widetilde{\xi}_{g}$ which is the  the intersection between the straight line $y^{0} + y^{1}=1$ and the straight line connecting the origin $(0\,,0)$ with the $\xi_{g}$.
Then, as long as $det(\xi)=c>0$, we see that for every $\xi_{g}$ there is one and only one such $\widetilde{\xi}_{g}$ (see figure \ref{Figure1}).
However, if $det(\xi)=c<0$, there will always be a point $\xi_{g}^{*}$ for which the straight line connecting it with the origin $(0\,,0)$ becomes parallel to the straight line $y^{0} + y^{1}=1$ (see figure \ref{Figure2}).
Consequently, $\widetilde{\xi}_{g}^{*}$ does not exist, and to divide $\xi_{g}=\mathbf{g}\,\xi\,\mathbf{g}^{\dagger}$ by its trace $Tr(\mathbf{g}\,\xi\,\mathbf{g}^{\dagger})$ is forbidden.

Below, we will introduce Hamiltonian and gradient-like vector fields on $\mathfrak{T}_{1}$ providing a realization of the Lie algebra $\mathfrak{sl}(\mathcal{H}\,,\mathbb{C})$ on $\mathfrak{T}_{1}$.
The action of these vector fields will integrate to an action of $SL(\mathcal{H}\,,\mathbb{C})$ only on the positive (negative) elements of $\mathfrak{T}_{1}$.
There, the fact that there are $\xi_{g}$ for which $\widetilde{\xi}_{g}$ does not exist is reflected in the fact that the gradient-like vector fields are non complete vector fields.
\end{rem}

By a slight modification of the arguments in \cite{grabowski_kus_marmo-geometry_of_quantum_systems_density_states_and_entanglement}
, it is possible to prove that $\stsp$ is partitioned into the disjoint union of orbits of $\alpha$.
Specifically:

\be
\stsp=\sqcup_{k=1}^{n}\,\stsp_{k}\,,
\ee
where $n=dim(\mathcal{H})$ and $\stsp_{k}$ denotes the space of positive matrices in $\mathfrak{T}_{1}$ having rank $k$, that is, the quantum states of rank $k$.
Every $\stsp_{k}$ possesses the structure of a differential manifold because they are homogeneous spaces of $SL(\mathcal{H}\,\mathbb{C})$, and thus, the standard tools of differential geometry applies.
Unfortunately, the whole $\stsp$ lacks of a differentiable structure except for $n=2$ \cite{grabowski_kus_marmo-geometry_of_quantum_systems_density_states_and_entanglement}.

To overcome this difficulty, we will be using the differential calculus on the space $\mathfrak{T}_{1}\subset\obsp^{*}$, which we think of as an ambient space for the space $\stsp$ of quantum states.

\vsp

We will now introduce some additional structures on $\mathfrak{T}_{1}\subset\obsp^{*}$ induced by the algebraic structures on  $\obsp$.
We start with the following definition:

\begin{defn}[Lie-Jordan algebra]\label{def: lie-jordan algebra}
Let $(A\,,\odot)$ denote a real Jordan algebra, and $(A\,,[[\,,]])$ a real Lie algebra.
Then $(A\,,\odot\,,[[\,,]])$ is called a Lie-Jordan algebra iff the following conditions hold: 
\begin{itemize}
\item $[[\mathbf{a}\,;\cdot]]$ is a derivation of $\odot$:

\be
\left[\left[\mathbf{a}\,,\mathbf{b}\odot \mathbf{c}\right]\right]]=\left[\left[[\mathbf{a}\,,\mathbf{b}\right]\right]\odot \mathbf{c} + \mathbf{b}\odot\left[\left[\mathbf{a}\,,\mathbf{c}\right]\right]\,;
\ee
\item the associator of $\odot$ is proportional to the Lie product:

\be\label{eqn: lo associatore del prodotto di Jordan risulta proporzionale al prodotto di Lie}
\left(\mathbf{a}\odot \mathbf{b}\right)\odot \mathbf{c} - \mathbf{a}\odot\left(\mathbf{b}\odot \mathbf{c}\right)=\left[\left[\mathbf{b}\,,\left[\left[\mathbf{c}\,,\mathbf{a}\right]\right]\,\right]\right]\,.
\ee
\end{itemize}
\end{defn}

The space $\obsp$ of self-adjoint elements in $\appa$ is naturally endowed with a Jordan  product $\odot$ and a Lie product $[[\,,]]$ given by:

\be
\mathbf{a}\odot\mathbf{b}:=\frac{(\mathbf{ab} + \mathbf{ba})}{2}\,,
\ee
\be
[[\mathbf{a}\,,\mathbf{b}]]:=\frac{\imath(\mathbf{ab} - \mathbf{ba})}{2}\,.
\ee
These product structures make $\obsp$ a Lie-Jordan algebra according to the previous definition.

Every element $\mathbf{a}\in\obsp$ may be represented as a linear function $f_{\mathbf{a}}$ on $\obsp^{*}$ as follows:

\be
f_{\mathbf{a}}(\xi):=\xi(\mathbf{a})\,.
\ee
Let\footnote{Throughout the rest of the paper, greek indexes will run from $0$ to $(n^{2}-1)$, while latin indexes will run from $1$ to $(n^{2}-1)$.} $\{\mathbf{e}^{\mu}\}_{\mu=0,...,n^{2}-1}$ be an orthonormal basis of $\obsp$, having $\mathbf{e}^{0}=\frac{\mathbb{I}}{\sqrt{n}}$. 
Then, we may define Cartesian coordinate system $\{x^{\mu}\}_{\mu=0,...,n^{2}-1}$ associated with $\{\mathbf{e}^{\mu}\}_{\mu=0,...,n^{2}-1}$ setting:

\be
x^{\mu}(\xi):=f_{\mathbf{e}^{\mu}}(\xi)=\xi(\mathbf{e}^{\mu})\,.
\ee
Defining:

\be\label{eqn: jordan on linear functions}
f_{\mathbf{a}}\odot f_{\mathbf{b}}:=f_{\mathbf{a}\odot\mathbf{b}}\,,
\ee

\be\label{eqn: lie on linear functions}
[[f_{\mathbf{a}}\,,f_{\mathbf{b}}]]:=f_{[[\mathbf{a}\,,\mathbf{b}]]}\,,
\ee
a direct computation shows that:

\begin{prop}
The set $(\mathcal{F}_{l}(\obsp^{*})\,,\odot\,,[[\,,]])$, where $\mathcal{F}_{l}(\obsp^{*})\subset\mathcal{F}(\obsp^{*})$ is the space of real linear functions on $\obsp^{*}$, and $\odot$, $[[\,,]]$ are given by (\ref{eqn: jordan on linear functions}) and (\ref{eqn: lie on linear functions}), provides a realization of the Lie-Jordan algebra $(\obsp\,,\odot\,,[[\,,]])$.
\end{prop}

Since the differentials of the linear functions generate the cotangent space at each point of $\obsp^{*}$, we can extend by linearity these two products $\odot$ and $[[\,, ]]$,  obtaining two contravariant tensor fields:

\be\label{eqn: G}
G= d^{\mu\nu}_{\sigma}x^{\sigma}\,\frac{\partial}{\partial x^{\mu}}\,\otimes\,\frac{\partial}{\partial x^{\nu}}\,.
\ee

\be\label{eqn: tildelambda}
\widetilde{\Lambda} = c^{\mu\nu}_{\sigma}x^{\sigma}\,\frac{\partial}{\partial x^{\mu}}\,\wedge\,\frac{\partial}{\partial x^{\nu}}\,,
\ee
The coefficients $c^{\mu\nu}_{\sigma}$ are the structure constants of the Lie product $[[\,,]]$ in $\obsp$.
Note that $2c^{\mu\nu}_{\sigma}$ are the structure constants of the Lie algebra $\mathfrak{u}(\mathcal{H})$ of the unitary group $\mathcal{U}(\mathcal{H})$, and thus, they are  antisymmetric in all indices.
Analogously, the coefficients $d^{\mu\nu}_{\sigma}$ are   the structure constants of the Jordan product $\odot$ in $\obsp$.
The structure constants $d^{\mu\nu}_{\sigma}$ are  symmetric in $\mu,\nu$, and we have $d^{0 0}_{j}=0$ and $d^{\mu\nu}_{0}=\frac{\delta^{\mu\nu}}{\sqrt{n}}$.
Furthermore,  the structure constants are invariant with respect to unitary transformations, that is, the structure constants of the basis $\{\mathbf{e}^{\mu}\}$ equal those of the basis $\{\mathbf{e}'^{\mu}\}$, where $\mathbf{e}'^{\mu}=\mathbf{U}\,\mathbf{e}^{\mu}\,\mathbf{U}^{\dagger}$ with $\mathbf{U}\mathbf{U}^{\dagger}=\mathbb{I}$.

Let $f,g\in\mathcal{F}(\obsp^{*})$, and let $\widetilde{\Lambda}$ and $G$ be the tensor fields in equations (\ref{eqn: tildelambda}) and (\ref{eqn: G}).
We define the following bilinear, binary product structures among functions on $\obsp^{*}$:

\be
\left\langle f\,;g\right\rangle:=G\left(\mathrm{d}f\,;\mathrm{d}g\right)\,.
\ee

\be
\left\{f\,;g\right\}:=\widetilde{\Lambda}\left(\mathrm{d}f\,;\mathrm{d}g\right)\,,
\ee
The second product is a Poisson bracket, while the first one is commutative but it does not possess additional properties, unless we restrict  it to a properly chosen subspace of functions, namely, linear functions.
In that case, we recover the Jordan structure of equation (\ref{eqn: jordan on linear functions}).

\vsp

Next, we define gradient-like and Hamiltonian vector fields:

\begin{defn}[Gradient-like and Hamiltonian vector fields on $\obsp^{*}$]\label{Gradient-like and Hamiltonian vector fields on the dual}
Let $f$ be a smooth function on $\obsp^{*}$, and let $G$ and $\widetilde{\Lambda}$ be as in equation (\ref{eqn: G}) and (\ref{eqn: tildelambda}).
Then, the gradient-like vector field $\mathbb{Y}_{f}$ and the Hamiltonian vector field $\widetilde{\mathbb{X}}_{f}$ associated with $f$ are defined as:

\be\label{eqn: gradient-like vector fields}
\mathbb{Y}_{f}:=G(\mathrm{d}f\,,\cdot)\,,
\ee
\be\label{eqn: hamiltonian vector fields}
\widetilde{\mathbb{X}}_{f}:=\widetilde{\Lambda}(\mathrm{d}f\,,\cdot)\,.
\ee
For the sake of notational simplicity, we will write $\mathbb{Y}_{\mathbf{a}}$ and $\widetilde{\mathbb{X}}_{\mathbf{a}}$ for the gradient-like and the Hamiltonian vector field associated with the linear function $f_{\mathbf{a}}$, where $\mathbf{a}\in\obsp$.
\end{defn}

Interestingly enough, we have:

\begin{prop}
Let $\mathbf{a},\mathbf{b}\in\obsp$, then the associated gradient-like and Hamiltonian vector fields satisfy the following commutation relations:

\be\label{eqn: commutation relations of the general linear group on ustarh by means of hamiltonian and gradient-like vector fields}
[\widetilde{\mathbb{X}}_{\mathbf{a}}\,,\widetilde{\mathbb{X}}_{\mathbf{b}}]=\widetilde{\mathbb{X}}_{[[\mathbf{a}\,,\mathbf{b}]]}\;\;\;\;\;\;\;\;[\widetilde{\mathbb{X}}_{\mathbf{a}}\,,\mathbb{Y}_{\mathbf{b}}]=\mathbb{Y}_{[[\mathbf{a}\,,\mathbf{b}]]}\;\;\;\;\;\;\;\;[\mathbb{Y}_{\mathbf{a}}\,,\mathbb{Y}_{\mathbf{b}}]=-\widetilde{\mathbb{X}}_{[[\mathbf{a}\,,\mathbf{b}]]}\,.
\ee
This means that the gradient-like and Hamiltonian vector fields associated with linear functions close on the Lie algebra $\mathfrak{gl}(\mathcal{H}\,,\mathbb{C})$ of the general linear group $GL(\mathcal{H}\,,\mathbb{C})$.

\begin{pf}
Recall that the differentials of the linear functions generate the cotangent space at each point of $\obsp^{*}$.
We start with the following computation:

$$
[\widetilde{\mathbb{X}}_{\mathbf{a}}\,,\widetilde{\mathbb{X}}_{\mathbf{b}}](f_{\mathbf{c}})=\widetilde{\mathbb{X}}_{\mathbf{a}}(\widetilde{\mathbb{X}}_{\mathbf{b}}(f_{\mathbf{c}})) - \widetilde{\mathbb{X}}_{\mathbf{b}}(\widetilde{\mathbb{X}}_{\mathbf{a}}(f_{\mathbf{c}}))= 
$$
$$
=\widetilde{\mathbb{X}}_{\mathbf{a}}(f_{[[\mathbf{b}\,,\mathbf{c}]]}) - \widetilde{\mathbb{X}}_{\mathbf{b}}(f_{[[\mathbf{a}\,,\mathbf{c}]]})=f_{[[\mathbf{a}\,,[[\mathbf{b}\,,\mathbf{c}]]]]} - f_{[[\mathbf{b}\,,[[\mathbf{a}\,,\mathbf{c}]]]]}
$$
According to the Jacobi identity of the Lie product we have:

\be\label{eqn: double lie bracket}
[[\mathbf{a}\,,[[\mathbf{b}\,,\mathbf{c}]]\,]] - [[\mathbf{b}\,,[[\mathbf{a}\,,\mathbf{c}]]\,]]=[[\,[[\mathbf{a}\,,\mathbf{b}]]\,,\mathbf{c}]]\,,
\ee
and thus:
\be
[\widetilde{\mathbb{X}}_{\mathbf{a}}\,,\widetilde{\mathbb{X}}_{\mathbf{b}}](f_{\mathbf{c}})=f_{[[[[\mathbf{a}\,,\mathbf{b}]]\,,\mathbf{c}]]}=\widetilde{\mathbb{X}}_{[[\mathbf{a}\,,\mathbf{b}]]}(f_{\mathbf{c}})\,,
\ee
which means:

\be
[\widetilde{\mathbb{X}}_{\mathbf{a}}\,,\widetilde{\mathbb{X}}_{\mathbf{b}}]=\widetilde{\mathbb{X}}_{[[\mathbf{a}\,,\mathbf{b}]]}\,.
\ee
Next, we have:

$$
[\widetilde{\mathbb{X}}_{\mathbf{a}}\,,\mathbb{Y}_{\mathbf{b}}](f_{\mathbf{c}})=\widetilde{\mathbb{X}}_{\mathbf{a}}(\mathbb{Y}_{\mathbf{b}}(f_{\mathbf{c}})) - \mathbb{Y}_{\mathbf{b}}(\widetilde{\mathbb{X}}_{\mathbf{a}}(f_{\mathbf{c}}))=
$$
$$
=\widetilde{\mathbb{X}}_{\mathbf{a}}\left(f_{\mathbf{b}\odot\mathbf{c}}\right) - \mathbb{Y}_{\mathbf{b}}(f_{[[\mathbf{a}\,,\mathbf{c}]]})=f_{[[\mathbf{a}\,,\mathbf{b}\odot\mathbf{c}]]}  - f_{\mathbf{b}\odot[[\mathbf{a}\,,\mathbf{c}]]}
$$
Recalling that $[[\mathbf{a}\,,]]$ is a derivation of $\odot$ for all $\mathbf{a}\in\obsp$, we have:

\be
[[\mathbf{a}\,,\mathbf{b}\odot\mathbf{c}]] - \mathbf{b}\odot[[\mathbf{a}\,,\mathbf{c}]]=[[\mathbf{a}\,,\mathbf{b}]]\odot\mathbf{c}\,,
\ee
and thus:

\be
[\widetilde{\mathbb{X}}_{\mathbf{a}}\,,\mathbb{Y}_{\mathbf{b}}](f_{\mathbf{c}})=f_{[[\mathbf{a}\,,\mathbf{b}]]\odot\mathbf{c}}\,,
\ee
which means:

\be
[\widetilde{\mathbb{X}}_{\mathbf{a}}\,,\mathbb{Y}_{\mathbf{b}}]=\mathbb{Y}_{[[\mathbf{a}\,,\mathbf{b}]]}\,.
\ee
Finally:

$$
[\mathbb{Y}_{\mathbf{a}}\,,\mathbb{Y}_{\mathbf{b}}](f_{\mathbf{c}})=\mathbb{Y}_{\mathbf{a}}(\mathbb{Y}_{\mathbf{b}}(f_{\mathbf{c}})) - \mathbb{Y}_{\mathbf{b}}(\mathbb{Y}_{\mathbf{a}}(f_{\mathbf{c}}))=
$$
$$
=\mathbb{Y}_{\mathbf{a}}\left(f_{\mathbf{b}\odot\mathbf{c}}\right) - \mathbb{Y}_{\mathbf{b}}(f_{\mathbf{a}\odot\mathbf{c}})=f_{\mathbf{a}\odot(\mathbf{b}\odot\mathbf{c})} - f_{\mathbf{b}\odot(\mathbf{a}\odot\mathbf{c})}\,.
$$
Using equations (\ref{eqn: lo associatore del prodotto di Jordan risulta proporzionale al prodotto di Lie}) and (\ref{eqn: double lie bracket}) we get:

\be\label{eqn: importante}
\mathbf{a}\odot(\mathbf{b}\odot\mathbf{c}) - \mathbf{b}\odot(\mathbf{a}\odot\mathbf{c})=-\left[\left[\,[[\mathbf{a}\,,\mathbf{b}]]\,\mathbf{c}\right]\right]\,,
\ee
and thus:

\be
[\mathbb{Y}_{\mathbf{a}}\,,\mathbb{Y}_{\mathbf{b}}](f_{\mathbf{c}})=-f_{\left[\left[\,[[\mathbf{a}\,,\mathbf{b}]]\,\mathbf{c}\right]\right]}\,,
\ee
which means

\be
[\mathbb{Y}_{\mathbf{a}}\,,\mathbb{Y}_{\mathbf{b}}]=-\widetilde{\mathbb{X}}_{[[\mathbf{a}\,,\mathbf{b}]]}\,.
\ee
Collecting the results, we have:

\be
[\widetilde{\mathbb{X}}_{\mathbf{a}}\,,\widetilde{\mathbb{X}}_{\mathbf{b}}]=\widetilde{\mathbb{X}}_{[[\mathbf{a}\,,\mathbf{b}]]}\;\;\;\;\;\;\;\;[\widetilde{\mathbb{X}}_{\mathbf{a}}\,,\mathbb{Y}_{\mathbf{b}}]=\mathbb{Y}_{[[\mathbf{a}\,,\mathbf{b}]]}\;\;\;\;\;\;\;\;[\mathbb{Y}_{\mathbf{a}}\,,\mathbb{Y}_{\mathbf{b}}]=-\widetilde{\mathbb{X}}_{[[\mathbf{a}\,,\mathbf{b}]]}\,,
\ee
which means that the Hamiltonian and gradient-like vector fields associated with linear functions close a representation of the Lie algebra $\mathfrak{gl}(\mathcal{H}\,,\mathbb{C})$ of the general linear group $GL(\mathcal{H}\,,\mathbb{C})$ on $\obsp^{*}$ as claimed.
\end{pf}
\end{prop}
A direct calculation shows that these vector fields generate the linear action 

$$
\xi\mapsto \widetilde{\alpha}(\mathbf{g}\,,\xi)=\mathbf{g}\,\xi\,\mathbf{g}^{\dagger}\,.
$$
The integral curves of the Hamiltonian vector field $\widetilde{\mathbb{X}}_{\mathbf{a}}$ are given by

$$
\gamma_{t}(\xi)=\mathbf{U}_{t}\,\xi\,\mathbf{U}_{t}^{\dagger}\,,
$$
with $\mathbf{U}_{t}=\exp(\imath t\mathbf{a})$, therefore, the integral curves starting at $\rho\in\stsp$ remain in $\stsp$ for all $t\in\mathbb{R}$. On the other hand, the integral curves of the gradient vector field $\mathbb{Y}_{\mathbf{a}}$ are

$$
\gamma_{t}(\xi)=\mathbf{A}_{t}\,\xi\,\mathbf{A}_{t}^{\dagger}\,,
$$
with $\mathbf{A}_{t}=\exp(t\mathbf{a})$, and thus the integral curves starting at $\rho\in\stsp$ exit from $\stsp$ because the trace is not preserved.

\vsp

\subsection*{Reduction}

We will now perform a reduction of the product structures $\langle\,,\rangle$ and $\{\,,\}$, as well as of the bivector fields $G$ and $\widetilde{\Lambda}$.
Then, we will define gradient-like and Hamiltonian vector fields on $\mathfrak{T}_{1}$, and it will turn out that these vector fields close on a realization of the Lie algebra $\mathfrak{sl}(\mathcal{H}\,,\mathbb{C})$ generating the nonlinear action $\alpha$ of equation (\ref{eq: nonlinear action of SL(n,C)}) on the space $\stsp$ of quantum states.

\vsp

In order to perform the reduction, let us briefly recall how product structures may pass to a quotient space.
First of all, let us consider a vector space $V$, and a closed linear subspace $W\subset V$.
It is well-known that the quotient space $E\equiv V/W$ inherits the structure of a vector space:

\be
[v_{1}] + [v_{2}]:= [v_{1} + v_{2}]\,.
\ee
If $V$ is the vector space $\mathcal{F}(M)$ of smooth functions on a differential manifold $M$, and $W$ is the closed subspace $\mathcal{I}_{\Sigma}$ of smooth functions vanishing on a submanifold $\Sigma\subset M$, we obtain $E\equiv \mathcal{F}(M)/\mathcal{I}_{\Sigma}\cong \mathcal{F}(\Sigma)$ \cite{carinena_ibort_marmo_morandi-geometry_from_dynamics_classical_and_quantum}. 
In particular, we have that the set $\mathfrak{T}_{1}$ is an affine subspace of $\obsp^{*}$, and thus its algebra of smooth functions can be identified with the quotient algebra $\mathcal{F}(\obsp^{*})/\mathcal{I}_{\mathfrak{T}_{1}}$, where $\mathcal{F}(\obsp^{*})$ is the algebra of smooth functions on $\obsp^{*}$, and $\mathcal{I}_{\mathfrak{T}_{1}}\subset \mathcal{F}(\obsp^{*})$ is the closed linear subspace consisting of smooth functions vanishing on $\mathfrak{T}_{1}$.

Now, let us endow the vector space $V$ with a product structure $\cdot$ compatible with $+$ so that $V$ becomes an algebra $A\equiv(V\,,+\,,\cdot)$.
It is clear that $W$ is again a closed linear subspace of $A$, however, as it stands it carries no information on the algebra structure of $A$.
This means that, in general, $E\equiv A/W$ will not be an algebra. 
If we want $E\equiv A/W$ to inherit an algebra structure, we must select $W$ so that it is an ideal of $A$.
In this case we can define the following product structure on $E$:

\be
[v_{1}]\cdot_{E}[v_{2}]:=[v_{1}\cdot v_{2}]\,.
\ee
Indeed, expressing $[v_{1}]$ and $[v_{2}]$ as the sum of a representative of the equivalence class with a generic element in $W$ we have:

\be
(v_{1} + w_{1})\cdot(v_{2} + w_{2})=v_{1}\cdot v_{2} + v_{1}\cdot w_{2} + w_{1}\cdot v_{2} + w_{1}\cdot w_{2}\equiv v_{1}\cdot v_{2} + w_{12}=[v_{1}\cdot v_{2}]\,,
\ee
where $w_{12}=v_{1}\cdot w_{2} + w_{1}\cdot v_{2} + w_{1}\cdot w_{2}$ is in $W$ for all $v_{1}$ and $v_{2}$ if and only if $W$ is an ideal of $\mathcal{A}$.

\vsp

Now, let us consider the algebra $(\mathcal{F}(\obsp^{*})\,,+\,,\{\,\})$. 
It is a matter of straightforward calculation to show that $\mathcal{I}_{\mathfrak{T}_{1}}$ is an ideal with respect to the product structure $\{\,,\}$, that is, $\{f\,,g\}\in\mathcal{I}_{\mathfrak{T}_{1}}$ whenever $f$ is in $\mathcal{I}_{\mathfrak{T}_{1}}$.
This means that on the quotient space $\mathcal{F}(\obsp^{*})/\mathcal{I}_{\mathfrak{T}_{1}}\cong\mathcal{F}(\mathfrak{T}_{1})$ we have the product structure $\{\,\}_{1}$:

\be
\{[f]\,,[g]\}_{1}:=\left[\{f\,,g\}\right]\,.
\ee
We can now use the product structure $\{\,,\}_{1}$ to construct a contravariant bivector field on $\mathfrak{T}_{1}$.
In order to do so, we identify the elements of the quotient space $\mathcal{F}(\obsp^{*})/\mathcal{I}_{\mathfrak{T}_{1}}$ with functions in $\mathcal{F}(\mathfrak{T}_{1})$.
If $\{x^{\mu}\}_{\mu=0,...,n^{2}-1}$ is the Cartesian coordinates in $\obsp^{*}$ associated with the orthonormal basis $\{\mathbf{e}^{\mu}\}_{\mu=0,...,n^{2}-1}$ introduced before, then the affine subspace $\mathfrak{T}_{1}$ may be identified with all those elements in $\obsp^{*}$ having $x^{0}=\frac{1}{\sqrt{n}}$.

\begin{rem}
What we have done here, is to select an origin in the affine subspace $\mathfrak{T}_{1}$, namely,  the point $\xi$ such that $x^{0}(\xi)=\frac{1}{\sqrt{n}}$ and $x^{j}(\xi)=0$ for all $j\neq0$. Interestingly, this point corresponds to the maximally mixed state.
\end{rem}

Denoting with $i\colon \mathfrak{T}_{1}\rightarrow\obsp^{*}$ the canonical immersion, we note that the pullback $f=\frac{A_{0}}{\sqrt{n}} + A_{j}\,x^{j}$ of a linear function $\widetilde{f}=A_{\mu}\,x^{\mu}\in\mathcal{F}(\obsp^{*})$ by means of $i$ is an affine function on $\mathfrak{T}_{1}$.
Consequently, we can select $(n^{2}-1)$ of them, say $f^{j}=x^{j}$, such that their differentials form a basis of the cotangent space $T^{*}_{\xi}\,\mathfrak{T}_{1}$ at each point $\xi\in\mathfrak{T}_{1}$.
An explicit calculation shows that:

\be
\{f_{j}\,,f_{k}\}_{1}=c_{jk}^{l}\,f_{l}\,.
\ee
Since $\{\mathrm{d}f_{j}\}_{j=1,...,n^{2}-1}$ is a basis of the cotangent space, we can define a contravariant bivector field $\Lambda$ setting:

\be
\Lambda(\mathrm{d}f_{j}\,,\mathrm{d}f_{k}):=\{f_{j}\,,f_{k}\}_{1}=c_{jk}^{l}\,f_{l}
\ee
and extending it by linearity.
The explicit expression of $\Lambda$ in the coordinate system associated with $\{f_{j}\}_{j=1,...,n^{2}-1}$ is:

\be\label{eqn: lambda on trace 1}
\Lambda=c^{jk}_{l}\,x^{l}\,\frac{\partial}{\partial x^{j}}\wedge\frac{\partial}{\partial x^{k}}\,.
\ee

\vsp
When we try to proceed similarly for $G$, we immediately find that $\mathcal{I}_{\mathfrak{T}_{1}}$ is not an ideal for the product structure $\langle\,,\rangle$ induced by $G$.
Our proposal to deal with this situation is to modify $G$ so that $\mathcal{I}_{\mathfrak{T}_{1}}$ becomes an ideal for the product structure associated with the new tensor field.
In doing so, we will lose the Jordan-Lie structure on the linear functions on $\obsp^{*}$,
Indeed, when we modify $G$, the resulting product will no longer be a Jordan product on linear functions, nor will it be compatible with the antisymmetric product $\{\,,\}$ associated with $\widetilde{\Lambda}$.
At the moment, we do not worry of this instance because we are primarily interested in defining  gradient-like and Hamiltonian vector fields generating the nonlinear action $\alpha$ in (\ref{eq: nonlinear action of SL(n,C)}).
A simple calculation shows that $\mathcal{I}_{\mathfrak{T}_{1}}$ is an ideal for the product structure associated with the contravariant tensor field:

\be
\widetilde{\mathcal{R}}:= G - \widetilde{\Delta}\otimes\widetilde{\Delta}  =d^{\mu\nu}_{\sigma}\,x^{\sigma}\,\frac{\partial}{\partial x^{\mu}}\otimes\frac{\partial}{\partial x^{\nu}} -  \widetilde{\Delta}\otimes\widetilde{\Delta}\,,
\ee
where $\widetilde{\Delta}=x^{\mu}\,\frac{\partial}{\partial x^{\mu}}$ is the Euler vector field representing the linear structure of $\obsp^{*}$.
The gradient-like vector fields associated with $\widetilde{\mathcal{R}}$ are:

\be\label{eqn: gradient-like vector fields on the self-adjoint part of the dual of the algebra}
\widetilde{\mathbb{Y}}_{f}:=\widetilde{\mathcal{R}}\left(\mathrm{d}f\,,\cdot\right)=G\left(\mathrm{d}f\,,\cdot\right) -  \widetilde{\Delta}(f)\,\widetilde{\Delta}=\mathbb{Y}_{f} -  \widetilde{\Delta}(f)  \,\widetilde{\Delta}\,.
\ee

Now, we can proceed in complete analogy with what has been done for $\{\,\}$ and $\widetilde{\Lambda}$.
The final result is the following symmetric contravariant tensor field:

\be\label{eqn: R on trace 1}
\mathcal{R}=\left(d^{jk}_{l}\,x^{l} + \frac{\delta^{jk}}{n}\right)\,\frac{\partial}{\partial x^{j}}\otimes\frac{\partial}{\partial x^{k}} -   \Delta\otimes\Delta
\ee  
where $\Delta=x^{j}\,\frac{\partial}{\partial x^{k}}$.
The symmetric product associated with $\mathcal{R}$ will be denoted as $\langle\,,\rangle_{1}$.

\vsp

Now that we have $\Lambda$ and $\mathcal{R}$, we can proceed and define gradient-like and Hamiltonian vector fields on $\mathfrak{T}_{1}$ in analogy with definition \ref{Gradient-like and Hamiltonian vector fields on the dual}:

\begin{defn}[Gradient-like and Hamiltonian vector fields on $\mathfrak{T}_{1}$]\label{Gradient-like and Hamiltonian vector fields on the trace one}
Let $f\in\mathcal{F}(\mathfrak{T}_{1})$, and $\mathcal{R}$ and $\Lambda$ be as in equation (\ref{eqn: R on trace 1}) and (\ref{eqn: lambda on trace 1}).
Then, the gradient-like vector field $Y_{f}$ and the Hamiltonian vector field $X_{f}$ associated with $f$ are defined as:

\be
Y_{f}:=\mathcal{R}\left(\mathrm{d}f\,;\cdot\right)\,,
\ee

\be
X_{f}:=\Lambda\left(\mathrm{d}f\,;\cdot\right)\,.
\ee
For the sake of notational simplicity, we will write $Y_{\mathbf{a}}$ and $X_{\mathbf{a}}$ for the gradient-like and the Hamiltonian vector field associated with the affine function $f_{\mathbf{a}}$, where $\mathbf{a}\in\obsp$.
\end{defn}

Writing $\xi=\frac{1}{\sqrt{n}}\mathbf{e}_{0} + x^{j}\mathbf{e}_{j}$, the explicit expressions of the gradient-like and Hamiltonian vector fields associated with the affine function $f_{\mathbf{a}}=\frac{a_{0}}{\sqrt{n}} + a_{j}x^{j}$, where $\mathbf{a}\in\obsp$, are:

\be\label{eqn: gradient-like vector field on trace 1}
Y_{\mathbf{a}}=Tr\left(\mathbf{a}\odot\xi\,\mathbf{e}^{k}\right)\,\frac{\partial}{\partial x^{k}} - f_{\mathbf{a}}\Delta =\left(d^{jk}_{l}\,x^{l}a_{j} +   \frac{\delta^{jk}a_{j}}{n} \right)\,\frac{\partial}{\partial x^{k}} -  x^{j}a_{j} \,\Delta\,,
\ee

\be\label{eqn: hamiltonian vector field on trace 1}
X_{\mathbf{a}}=c^{jk}_{l}\,x^{l}a_{j}\,\frac{\partial}{\partial x^{k}}\,.
\ee
Note that the gradient-like vector fields contain a quadratic term with respect to the coordinate system $\{x^{k}\}_{k=1,...,n^{2}-1}$ adapted to $\mathfrak{T}_{1}$.
Moreover, note that the $0$-component of $\mathbf{a}$ does not play any role in the definition of $X_{\mathbf{a}}$ and $Y_{\mathbf{a}}$.
In particular, the Hamiltonian and gradient-like vector fields associated with $f_{\mathbf{a}}$ are  everywhere vanishing  whenever $\mathbf{a}=a_{0}\mathbf{e}^{0}$.

There is a very interesting relation between the vector fields $\widetilde{\mathbb{X}}_{\mathbf{a}}$, $\widetilde{\mathbb{Y}}_{\mathbf{a}}$, $X_{\mathbf{a}}$ and $Y_{\mathbf{a}}$.
To see this, recall that the vector fields on $\obsp^{*}$ are derivations of the pointwise product of $\mathcal{F}(\obsp^{*})$.
Any such derivation, say $\widetilde{D}$, defines a derivation $D$ of the quotient algebra (with respect to the  pointwise product) if and only if $D(\mathcal{I}_{\mathfrak{T}_{1}})\subset \mathcal{I}_{\mathfrak{T}_{1}}$ \cite{carinena_ibort_marmo_morandi-geometry_from_dynamics_classical_and_quantum}, indeed, we can define:

\be
D([f]):=[\widetilde{D}(f)]\,.
\ee
It is then clear that Hamiltonian vector fields $\widetilde{\mathbb{X}}_{\mathbf{a}}$ and gradient-like vector fields $\widetilde{\mathbb{Y}}_{\mathbf{a}}$ define derivations of $\mathcal{F}(\obsp^{*})/\mathcal{I}_{\mathfrak{T}_{1}}$.
Once we identify $\mathcal{F}(\obsp^{*})/\mathcal{I}_{\mathfrak{T}_{1}}$ with $\mathcal{F}(\mathfrak{T}_{1})$, it is possible to show that the derivation associated with $\widetilde{\mathbb{X}}_{\mathbf{a}}$ is the Hamiltonian vector field $X_{\mathbf{a}}$ associated with $\mathbf{a}$ by means of $\Lambda$.
Similarly, the derivation associated with $\widetilde{\mathbb{Y}}_{\mathbf{a}}$ is the gradient-like vector field $Y_{\mathbf{a}}$ associated with $\mathbf{a}$ by means of $\mathcal{R}$.

\vsp

We now have the following result:

\begin{prop}
Let $\mathbf{a},\mathbf{b}\in\obsp$ be such that $Tr(\mathbf{a})=Tr(\mathbf{b})=0$.
Then the associated gradient-like and Hamiltonian vector fields on $\mathfrak{T}_{1}$ satisfy the following commutation relations:

\be\label{eqn: commutation relations on trace 1 hamiltonian and gradient-like vector fields}
[X_{\mathbf{a}}\,,X_{\mathbf{b}}]=X_{[[\mathbf{a}\,,\mathbf{b}]]}\;\;\;\;\;\;\;\;[X_{\mathbf{a}}\,,Y_{\mathbf{b}}]=Y_{[[\mathbf{a}\,,\mathbf{b}]]}\;\;\;\;\;\;\;\;[Y_{\mathbf{a}}\,,Y_{\mathbf{b}}]=-X_{[[\mathbf{a}\,,\mathbf{b}]]}\,.
\ee
This means that the gradient-like and Hamiltonian vector fields associated with affine functions close on the Lie algebra $\mathfrak{sl}(\mathcal{H}\,,\mathbb{C})$ of the special linear group $SL(\mathcal{H}\,,\mathbb{C})$.

\begin{pf}
Since the affine functions $f_{\mathbf{a}}$ are enough to generate the cotangent space at each point, we will compute the commutators evaluating them on the affine functions themselves.
For the Hamiltonian vector fields we have:

$$
[X_{\mathbf{a}}\,,X_{\mathbf{b}}](f_{\mathbf{c}})=X_{\mathbf{a}}(X_{\mathbf{b}}(f_{\mathbf{c}})) - X_{\mathbf{b}}(X_{\mathbf{a}}(f_{\mathbf{c}}))= 
$$
$$
=X_{\mathbf{a}}(f_{[[\mathbf{b}\,,\mathbf{c}]]}) - X_{\mathbf{b}}(f_{[[\mathbf{a}\,,\mathbf{c}]]})=f_{[[\mathbf{a}\,,[[\mathbf{b}\,,\mathbf{c}]]]]} - f_{[[\mathbf{b}\,,[[\mathbf{a}\,,\mathbf{c}]]]]}\,,
$$
where we have used:

\be
X_{\mathbf{a}}(f_{\mathbf{b}})=\Lambda(\mathrm{d}f_{\mathbf{a}}\,,\mathrm{d}f_{\mathbf{b}})=\{f_{\mathbf{a}}\,,f_{\mathbf{b}}\}_{1}=f_{[[\mathbf{a}\,,\mathbf{b}]]}\,.
\ee
It is easy to see that:

\be
[[\mathbf{a}\,,[[\mathbf{b}\,,\mathbf{c}]]\,]] - [[\mathbf{b}\,,[[\mathbf{a}\,,\mathbf{c}]]\,]]=[[\,[[\mathbf{a}\,,\mathbf{b}]]\,,\mathbf{c}]]\,,
\ee
from which it follows that:
\be
[X_{\mathbf{a}}\,,X_{\mathbf{b}}](f_{\mathbf{c}})=f_{[[[[\mathbf{a}\,,\mathbf{b}]]\,,\mathbf{c}]]}=X_{[[\mathbf{a}\,,\mathbf{b}]]}(f_{\mathbf{c}})\,,
\ee
and thus:

\be
[X_{\mathbf{a}}\,,X_{\mathbf{b}}]=X_{[[\mathbf{a}\,,\mathbf{b}]]}\,.
\ee

Before computing the commutator between Hamiltonian and gradient-like vector fields, let us note that:

\be\label{eqn: gradient-like action on affine functions}
Y_{\mathbf{a}}(f_{\mathbf{b}})=\mathcal{R}(\mathrm{d}f_{\mathbf{a}}\,,\mathrm{d}f_{\mathbf{b}})=d^{jk}_{l}a_{j}b_{k} + \frac{\delta^{jk}a_{j}b_{k}}{n} -  x^{j}a_{j}\,x^{k}b_{k} \,.
\ee
Now, the Jordan product $\mathbf{a}\odot\mathbf{b}$ reads:

\be\label{eqn: explicit jordan product}
\mathbf{a}\odot\mathbf{b}=d^{\mu\nu}_{\sigma}a_{\mu}b_{\nu}\,\mathbf{e}^{\sigma}=d^{jk}_{0}a_{j}b_{k}\,\mathbf{e}^{0} + d^{jk}_{l}a_{j}b_{k}\,\mathbf{e}^{l}=\frac{\delta^{jk}a_{j}b_{k}}{\sqrt{n}}\,\mathbf{e}^{0} + d^{jk}_{l}a_{j}b_{k}\,\mathbf{e}^{l}\,,
\ee
where we used $d^{0 0}_{j}=0$, $d^{\mu\nu}_{0}=\frac{\delta^{\mu\nu}}{\sqrt{n}}$, and the fact that $\mathbf{a}$ and $\mathbf{b}$ are traceless.
Comparing equation (\ref{eqn: gradient-like action on affine functions}) with equation (\ref{eqn: explicit jordan product}) it follows that 
\be
Y_{\mathbf{a}}(f_{\mathbf{b}})=f_{\mathbf{a}\odot\mathbf{b}}  - f_{\mathbf{a}}f_{\mathbf{b}} \,.
\ee
Computing the commutator, we have:

$$
[X_{\mathbf{a}}\,,Y_{\mathbf{b}}](f_{\mathbf{c}})=X_{\mathbf{a}}(Y_{\mathbf{b}}(f_{\mathbf{c}})) - Y_{\mathbf{b}}(X_{\mathbf{a}}(f_{\mathbf{c}}))=X_{\mathbf{a}}\left(f_{\mathbf{b}\odot\mathbf{c}}-  f_{\mathbf{b}}f_{\mathbf{c}} \right) - Y_{\mathbf{b}}(f_{[[\mathbf{a}\,,\mathbf{c}]]})=
$$
$$
=f_{[[\mathbf{a}\,,\mathbf{b}\odot\mathbf{c}]]} - f_{\mathbf{c}}\,f_{[[\mathbf{a}\,,\mathbf{b}]]}  -  f_{\mathbf{b}}\,f_{[[\mathbf{a}\,,\mathbf{c}]]}  - f_{\mathbf{b}\odot[[\mathbf{a}\,,\mathbf{c}]]} + \ f_{\mathbf{b}}f_{[[\mathbf{a}\,,\mathbf{c}]]} =
$$
$$
=f_{[[\mathbf{a}\,,\mathbf{b}\odot\mathbf{c}]]} - f_{\mathbf{c}}\,f_{[[\mathbf{a}\,,\mathbf{b}]]} - f_{\mathbf{b}\odot[[\mathbf{a}\,,\mathbf{c}]]} \,.
$$
A direct computation shows that:

\be
[[\mathbf{a}\,,\mathbf{b}\odot\mathbf{c}]] - \mathbf{b}\odot[[\mathbf{a}\,,\mathbf{c}]]=[[\mathbf{a}\,,\mathbf{b}]]\odot\mathbf{c}\,,
\ee
and thus:

\be
[X_{\mathbf{a}}\,,Y_{\mathbf{b}}](f_{\mathbf{c}})=f_{[[\mathbf{a}\,,\mathbf{b}]]\odot\mathbf{c}}  - f_{[[\mathbf{a}\,,\mathbf{b}]]}\,f_{\mathbf{c}} \,,
\ee
which means:

\be
[X_{\mathbf{a}}\,,Y_{\mathbf{b}}]=Y_{[[\mathbf{a}\,,\mathbf{b}]]}\,.
\ee
Finally, noting that:

$$
Y_{\mathbf{a}}\left(f_{\mathbf{b}\odot\mathbf{c}}  -  f_{\mathbf{b}}f_{\mathbf{c}} \right)=f_{\mathbf{a}\odot(\mathbf{b}\odot\mathbf{c})}  -  f_{\mathbf{a}}f_{\mathbf{b}\odot\mathbf{c}}  - f_{\mathbf{c}}\left(f_{\mathbf{a}\odot\mathbf{b}}  - f_{\mathbf{a}}f_{\mathbf{b}} \right) - f_{\mathbf{b}}\left(f_{\mathbf{a}\odot\mathbf{c}} -  f_{\mathbf{a}}f_{\mathbf{c}} \right)
$$
we have:

$$
[Y_{\mathbf{a}}\,,Y_{\mathbf{b}}](f_{\mathbf{c}})=Y_{\mathbf{a}}(Y_{\mathbf{b}}(f_{\mathbf{c}})) - Y_{\mathbf{b}}(Y_{\mathbf{a}}(f_{\mathbf{c}}))=Y_{\mathbf{a}}\left(f_{\mathbf{b}\odot\mathbf{c}}  -  f_{\mathbf{b}}f_{\mathbf{c}} \right) - Y_{\mathbf{b}}\left(f_{\mathbf{a}\odot\mathbf{c}} -  f_{\mathbf{a}}f_{\mathbf{c}} \right)=
$$

$$
=f_{\mathbf{a}\odot(\mathbf{b}\odot\mathbf{c})} -  f_{\mathbf{a}}f_{\mathbf{b}\odot\mathbf{c}}  - 	 f_{\mathbf{c}} \left(f_{\mathbf{a}\odot\mathbf{b}}  -  f_{\mathbf{a}}f_{\mathbf{b}} \right) -  f_{\mathbf{b}} \left(f_{\mathbf{a}\odot\mathbf{c}}  -  f_{\mathbf{a}}f_{\mathbf{c}} \right) - 
$$
$$
-f_{\mathbf{b}\odot(\mathbf{a}\odot\mathbf{c})}  +  f_{\mathbf{b}}f_{\mathbf{a}\odot\mathbf{c}}  +  f_{\mathbf{c}} \left(f_{\mathbf{a}\odot\mathbf{b}}  -  f_{\mathbf{a}}f_{\mathbf{b}} \right) +  f_{\mathbf{a}} \left(f_{\mathbf{b}\odot\mathbf{c}}  -  f_{\mathbf{b}}f_{\mathbf{c}} \right)=-f_{\left[\left[\,[[\mathbf{a}\,,\mathbf{b}]]\,\mathbf{c}\right]\right]} 
$$
where, in the last equality, we used equation (\ref{eqn: importante}).
Eventually, we get:

\be
[Y_{\mathbf{a}}\,,Y_{\mathbf{b}}]=-X_{[[\mathbf{a}\,,\mathbf{b}]]}\,.
\ee
Collecting the results we have:

\be
[X_{\mathbf{a}}\,,X_{\mathbf{b}}]=X_{[[\mathbf{a}\,,\mathbf{b}]]}\;\;\;\;\;\;\;\;[X_{\mathbf{a}}\,,Y_{\mathbf{b}}]=Y_{[[\mathbf{a}\,,\mathbf{b}]]}\;\;\;\;\;\;\;\;[Y_{\mathbf{a}}\,,Y_{\mathbf{b}}]=-X_{[[\mathbf{a}\,,\mathbf{b}]]}\,,
\ee
which defines a realization of the Lie algebra $\mathfrak{sl}(\mathcal{H}\,,\mathbb{C})$ of the special linear group $SL(\mathcal{H}\,,\mathbb{C})$  as claimed.
\end{pf}
\end{prop}
One could be tempted to say that the realization of $\mathfrak{sl}(\mathcal{H}\,,\mathbb{C})$ by means of Hamiltonian and gradient-like vector fields associated with affine functions, integrates to an action of $SL(\mathcal{H}\,,\mathbb{C})$ on $\mathfrak{T}_{1}$ just as it happens for the representation of $\mathfrak{gl}(\mathcal{H}\,,\mathbb{C})$ on $\obsp^{*}$ (see equation (\ref{eqn: commutation relations of the general linear group on ustarh by means of hamiltonian and gradient-like vector fields})).
However, this is not the case.
What happens is that the gradient-like vector fields in equation (\ref{eqn: commutation relations on trace 1 hamiltonian and gradient-like vector fields}) are, in general, not complete, and thus the Lie algebra realization does not integrate to an action of the Lie group.

What is very interesting though, is that on the positive elements of $\mathfrak{T}_{1}$, that is, the quantum states, these vector fields are complete, and their flow is precisely the action $\alpha$ of equation (\ref{eq: nonlinear action of SL(n,C)}):

\begin{prop}\label{prop: integration of nonlinear action on quantum states}
Let $\mathbf{a},\mathbf{b}\in\obsp$ be such that $Tr(\mathbf{a})=Tr(\mathbf{b})=0$.
Then, the integral curve of $X_{\mathbf{a}} + Y_{\mathbf{b}}$ starting at $\rho\in\stsp$ is given by:

\be
\gamma_{t}(\rho):=\alpha_{\mathbf{g}_{t}}(\rho)=\frac{\mathbf{g}_{t}\,\rho\,\mathbf{g}_{t}^{\dagger}}{Tr(\mathbf{g}_{t}\,\rho\,\mathbf{g}_{t}^{\dagger})}\,,
\ee
where $\mathbf{A}=\mathbf{a} + \imath\mathbf{b}\in\mathfrak{sl}(\mathcal{H}\,,\mathbb{C})$ so that $\mathbf{g}_{t}=\exp(\frac{t}{2}\mathbf{A})$ is in $SL(\mathcal{H}\,,\mathbb{C})$ for all $t$.

\begin{pf}
Let us start writing $\rho=\frac{1}{\sqrt{n}}\mathbf{e}_{0} + x^{l}\mathbf{e}_{l}$, and  compute the derivative of $\gamma_{t}$ with respect to $t$ in $t=0$:

$$
\left.\frac{\mathrm{d} \gamma_{t}(\rho)}{\mathrm{d} t}\right|_{t=0}=\frac{1}{2}\left(\mathbf{A}\,\rho + \rho\,\mathbf{A}^{\dagger} - \rho\,Tr\left(\mathbf{A}\,\rho + \rho\,\mathbf{A}^{\dagger}\right)\right)=
$$
$$
=[[\mathbf{b}\,,\rho]] + \mathbf{a}\odot\rho - Tr(\mathbf{a}\odot \rho)\rho =
$$
\be\label{eqn: derivative of nonlinear action}
= c^{jk}_{l}x^{l}b_{j}\,\mathbf{e}_{k} + d^{jk}_{l}x^{l}a_{j} \mathbf{e}_{k} +  \frac{\delta^{jk} a_{j}}{n} \,\mathbf{e}_{k} - \delta^{j}_{l}a_{j}x^{l} \,\rho^{k}\mathbf{e}_{k} \,.
\ee
This is the tangent vector of $\gamma_{t}$ at $\rho$ expressed in matrix form.
Its $k$-th component along the basis vector $\mathbf{e}^{k}$ is obtained taking  the trace of $\left.\frac{\mathrm{d} \gamma_{t}(\rho)}{\mathrm{d} t}\right|_{t=0}$ with $\mathbf{e}^{k}$:

\be\label{eqn: kth component of tangent vector of nonlinear action}
Tr\left(\left.\frac{\mathrm{d} \gamma_{t}(\rho)}{\mathrm{d} t}\right|_{t=0}\,\mathbf{e}^{k}\right)=c^{jk}_{l}x^{l}b_{j} + d^{jk}_{l}x^{l}a_{j} +  \frac{\delta^{jk} a_{j}}{n}  - \delta^{j}_{l}a_{j}x^{l} \,x^{k}\,.
\ee
The proposition follows from the comparison between equation (\ref{eqn: kth component of tangent vector of nonlinear action}) with equations (\ref
{eqn: gradient-like vector field on trace 1}) and  (\ref{eqn: hamiltonian vector field on trace 1}).
\end{pf}
\end{prop}
From proposition \ref{prop: integration of nonlinear action on quantum states} we conclude that $\gamma_{t}(\rho)$ is an integral curve for $Y_{\mathbf{a}} + X_{\mathbf{b}}$ with initial point $\rho$ for all  $\rho\in\stsp\subset\mathfrak{T}_{1}$.
These integral curves are clearly complete,  and they lie entirely in the space $\stsp$ of quantum states.
Specifically, if $\rho\in\stsp_{k}$ then $\gamma_{t}(\rho)$ is in $\stsp_{k}$ for all $t\in\mathbb{R}$.
If $\mathbf{A}=\imath\mathbf{b}$, then $\gamma_{t}(\rho)=\mathbf{U}_{t}\,\rho\,\mathbf{U}_{t}^{\dagger}$, and thus $\gamma_{t}(\rho)$ lies entirely in the set of isospectral states.
Consequently, $X_{\mathbf{b}}$ represents the vector field generating the unitary part of a quantum dynamical process. 
On the other hand, when $\mathbf{A}=\mathbf{a} + \imath\mathbf{b}$, with $\mathbf{a}\neq0$ and $\mathbf{b}$ arbitrary, the curve $\gamma_{t}(\rho)$ is generically transversal to the set of isospectral states, however remaining entirely in the set of quantum states with fixed rank.
Clearly, since $Y_{\mathbf{a}}$ contains a quadratic term, its integral curves can not represent linear quantum dynamical processes.
We will see that if we combine $Y_{\mathbf{a}}$ with a properly defined vector field $Z_{\mathcal{K}}$, then $Y_{\mathbf{a}} + Z_{\mathcal{K}}$ does not contain any quandratic term, and its integral curves represent the dissipative part of quantum dynamical processes.

\begin{exmp}[Two-level quantum system]\label{example: two level quantum system geometry}

To illustrate our general arguments we consider the example of a two-level quantum system.
To make contact with the widespread notation for qubit, we will here drop the requirement of orthonormality for the basis $\{\mathbf{e}_{\mu}\}_{\mu=0,...,3}$ and consider the orthogonal basis generated by the Pauli  matrices:

\be
\sigma_{0}=\left(\begin{array}{cc} 1 & 0 \\ 0 & 1 \end{array}\right) \;\;\;\;\;\;\;\;\;\; \sigma_{1}=\left(\begin{array}{cc} 0 & 1 \\ 1 & 0 \end{array}\right)
\ee 
\be
\sigma_{2}=\left(\begin{array}{cc} 0 & -\imath \\ \imath & 0 \end{array}\right) \;\;\;\;\;\;\;\;\;\; \sigma_{3}=\left(\begin{array}{cc} 1 & 0 \\ 0 & -1 \end{array}\right)\,.
\ee
This choice will affect some numerical factors in the coefficients $c^{jk}_{l}$ and $d^{\mu\nu}_{\alpha}$.
However, from the practical point of view, it is a convenient choice because of the peculiar properties of the Pauli matrices.
A quantum state $\rho$ is written as:

\be
\rho=\frac{1}{2}\left(\sigma_{0} + \mathbf{x}\cdot\boldsymbol{\sigma}\right) 
\ee
with $|\mathbf{x}|^{2}\leq 1$.
In this case, $\stsp$ has only two strata, namely, $\stsp_{1}$ and $\stsp_{2}$, and it is a proper manifold with boundary.
As shown in \cite{grabowski_kus_marmo-geometry_of_quantum_systems_density_states_and_entanglement} this is the only case in which $\stsp$ is a differential manifold with a smooth boundary.
Specifically, $\stsp$ is the $3$-dimensional solid ball and the two strata are the surface of the ball, that is, the pure states; and the open interior of the ball, that is, the mixed states.
It should be noticed that while pure states are represented by a compact manifold without boundary, the stratum of mixed states is bounded but not compact, and its closure is the whole space of quantum states $\stsp$.

The expressions for $\mathcal{R}$ and $\Lambda$ are:

\be
\mathcal{R}= \delta^{jk}\,\frac{\partial}{\partial x^{j}}\otimes\frac{\partial}{\partial x^{k}} - \Delta\otimes\Delta\,,
\ee
\be
\Lambda= -\epsilon^{jk}_{l}\,x^{l}\frac{\partial}{\partial x^{j}}\wedge\frac{\partial}{\partial x^{k}}\,,
\ee
where $\epsilon^{jk}_{l}$ is the Levi-Civita symbol.
Gradient vector fields associated with the affine function $f_{\sigma_{j}}$ are:

\be
Y_{j}=\frac{\partial}{\partial x^{j}} - x^{j}\,\Delta \,,
\ee
while Hamiltonian ones read:

\be
X_{j}=-\epsilon^{jk}_{l}\,x^{l}\,\frac{\partial}{\partial x^{k}}\,.
\ee
Together they close on the Lie algebra of $SL(2,\mathbb{C})$.
Furthermore, the Hamiltonian ones are tangent to the sphere of radius $r$ for all $r>0$:

\be
\mathcal{L}_{X_{j}}r^{2}=-\epsilon^{jk}_{l}\,x^{l}x^{k}=0\,,
\ee
while the gradient ones are tangent only to the sphere of radius $r=1$ (the pure states), in fact we get:

\be
\mathcal{L}_{Y_{j}}r^{2}=(1 - r^{2})x^{j}\,.
\ee

\end{exmp}

\section{GKLS vector field on $\stsp$}\label{sec:GKLS vector field}

As it was discussed in the introduction and  according to \cite{gorini_kossakowski_sudarshan-completely_positive_dynamical_semigroups_of_N-level_systems} and \cite{lindblad-on_the_generators_of_quantum_dynamical_semigroups},  the generator $\mathbf{L}$ of a linear quantum dynamical process can be expressed as a linear operator on $\xi$ as follows:

\be\label{eqn: K-L equation}
\mathbf{L}(\xi)=-2\left[\left[\mathbf{H}\,,\xi\right]\right] - \mathbf{V}\odot\xi +\mathcal{K}(\xi)\,,
\ee
where $\mathbf{H}\in\obsp$, $\mathbf{v}_{j}\in\mathcal{B}(\mathcal{H})$,  $\mathbf{V}=\sum_{j=1}^{N}\mathbf{v}_{j}^{\dagger}\mathbf{v}_{j}$, and the linear map $\mathcal{K}$ is a completely-positive map: 

\be
\mathcal{K}(\xi)= \sum_{j=1}^{N}\,\mathbf{v}_{j}\,\xi\, \mathbf{v}^{\dagger}_{j}\;\;\;\;\;\mbox{ with } \;N\leq(n^{2}-1)\,.
\ee
If $\mathcal{H}$ is finite-dimensional, this is the most general form for the generator of a dynamical process which is linear, completely positive and trace preserving (CPTP) \cite{gorini_kossakowski_sudarshan-completely_positive_dynamical_semigroups_of_N-level_systems,lindblad-on_the_generators_of_quantum_dynamical_semigroups}.




\vsp

The integration of the equations of motion associated with $\mathbf{L}$ gives a one-parameter semigroup $\{\Phi_{\tau}\}$ of completely-positive maps $\Phi_{\tau}:\stsp\rightarrow\stsp$ for $\tau\geq0$, such that $\Phi_{0}$ is the identity transformation.
Actually, $\{\Phi_{\tau}\}$ is well-defined and differentiable for all $\tau\in\mathbb{R}$ on the whole $\obsp^{*}$, but, for $\tau<0$ it fails to preserve positivity, hence, it maps quantum states out of $\stsp$.

We will now analyze the vector field $\widetilde{\Gamma}\equiv Z_{\mathbf{L}}$ associated with the GKLS generator $\mathbf{L}$.
For this purpose, let $\{\mathbf{e}_{\mu}\}_{\mu=0,...,n^{2}-1}$ denote the basis in $\obsp^{*}$ which is dual to the orthonormal basis $\{\mathbf{e}^{\mu}\}_{\mu=0,...,n^{2}-1}$ of $\obsp$ introduced before.

\begin{defn}[Linear vector field associated with a linear map]\label{def: linear vector field from linear map}
Let $A(\xi)=A^{\mu}_{\nu}\xi^{\nu}\,\mathbf{e}_{\mu}$ be a linear map from $\obsp^{*}$ to itself, and let $\{x^{\mu}\}$ be the Cartesian coordinates system associated with $\{\mathbf{e}^{\mu}\}_{\mu=0,...,n^{2}-1}$.
We define a linear vector field $\mathbb{Z}_{A}$ on $\obsp^{*}$ associated with $A$ as follows (see \cite{carinena_ibort_marmo_morandi-geometry_from_dynamics_classical_and_quantum} chapter $2$ and $3$):

\be\label{eqn: linear vector field from linear map}
\mathbb{Z}_{A}:=A^{\mu}_{\nu}\,x^{\nu}\,\frac{\partial}{\partial x^{\mu}}\,.
\ee
Its action on linear functions reads:

\be
Z_{A}(f_{\mathbf{b}})(\xi)=f_{\mathbf{b}}(A(\xi))=A^{\mu}_{\nu}\,b_{\mu}x^{\nu}\,.
\ee
\end{defn}

It is a matter of straightforward calculation to  prove that:

\begin{prop}\label{prop: linar combinations of linear maps give rise to linear combinations of linear vector fields}
Let $A,B$ be linear maps from $\obsp^{*}$ to itself, then:

\be
Z_{A+B}=Z_{A} + Z_{B}\,.
\ee
\end{prop}

The GKLS generator $\mathbf{L}$ is a linear map from $\obsp^{*}$ to itself, therefore, we may define its associated linear vector field $\widetilde{\Gamma}\equiv Z_{L}$ on $\obsp^{*}$ by means of definition \ref{def: linear vector field from linear map}.

\begin{prop}[GKLS vector field on $\obsp^{*}$]
Let $\mathbf{L}$ be the GKLS generator of equation (\ref{eqn: K-L equation}).
Then:

\be
\widetilde{\Gamma}= \widetilde{\mathbb{X}}_{\mathbf{a}} + \mathbb{Y}_{\mathbf{b}} + \mathbb{Z}_{\mathcal{K}}\,,
\ee
where $\widetilde{\mathbb{X}}_{\mathbf{a}}$ is the Hamiltonian vector field associated with  $\mathbf{a}= - 2\mathbf{H}$ by means of $\widetilde{\Lambda}$, the gradient-like vector field $\mathbb{Y}_{\mathbf{b}}$ is the one associated with $\mathbf{b}=-\mathbf{V}$ by means of $G$, and $\mathbb{Z}_{\mathcal{K}}$ is the linear vector field associated with the CPTP map $\mathcal{K}$ by means of (\ref{eqn: linear vector field from linear map}).
\begin{pf}
Let us start writing:

\be
\mathbf{L}(\xi)=2\left[\left[\mathbf{H}\,,\xi\right]\right] - \mathbf{V}\odot\xi +\mathcal{K}(\xi) \equiv -2C_{\mathbf{H}}(\xi) - A_{\mathbf{V}}(\xi) + \mathcal{K}(\xi)\,,
\ee
where the linear maps $C_{\mathbf{H}}$ and $A_{\mathbf{V}}$ are given by:

$$
C_{\mathbf{H}}(\xi):=[[\mathbf{H}\,,\xi]]=c^{\mu\nu}_{\sigma}h_{\nu}x^{\sigma}\,\mathbf{e}_{\mu}\,,
$$ 
and

$$
A_{\mathbf{V}}(\xi):=\mathbf{V}\odot\xi=d^{\mu\nu}_{\sigma}V_{\nu}x^{\sigma}\,\mathbf{e}_{\mu}\,.
$$
According to definition \ref{def: linear vector field from linear map} we have:

\be
Z_{C_{\mathbf{H}}}=c^{\mu\nu}_{\sigma}h_{\nu}x^{\sigma}\,\frac{\partial}{\partial x^{\mu}}=\widetilde{\mathbb{X}}_{\mathbf{H}}\,,
\ee

\be
Z_{A_{\mathbf{V}}}=d^{\mu\nu}_{\sigma}V_{\nu}x^{\sigma}\,\frac{\partial}{\partial x^{\mu}}=\mathbb{Y}_{\mathbf{V}}\,,
\ee
where we have used the coordinate expressions of Hamiltonian and gradient-like vector fields given by equations (\ref{eqn: hamiltonian vector field on trace 1}) and (\ref{eqn: gradient-like vector field on trace 1}).
Since $\mathbf{L}$ is a linear combination of the three linear maps $C_{\mathbf{H}}(\xi)$,   $A_{\mathbf{V}}(\xi)$, and $\mathcal{K}(\xi)$, the results follows from proposition \ref{prop: linar combinations of linear maps give rise to linear combinations of linear vector fields}\,.
\end{pf}
\end{prop}

\subsection{GKLS vector field on $\mathfrak{T}_{1}$}

Our aim is to define a vector field on $\mathfrak{T}_{1}$ representing the GKLS generator $\mathbf{L}$.
We will construct such a vector field by means of a reduction procedure applied to $\Gamma_{L}$.

\begin{prop}[GKLS vector field on $\mathfrak{T}_{1}$]
The linear vector field $\widetilde{\Gamma}$ defines a derivation, say $\Gamma$, of the algebra $\mathcal{F}(\mathfrak{T}_{1})$.
Furthermore, $\Gamma$ will decompose into the sum of three vector fields all of which, spearately, define derivations of $\mathcal{F}(\mathfrak{T}_{1})$.
Specifically, we have:

\be\label{eqn: GKLS vector field}
\Gamma = X_{\mathbf{a}} + Y_{\mathbf{b}} + Z_{\mathcal{K}}\,,
\ee
where $X_{\mathbf{a}}$ is the Hamiltonian vector field associated with $\mathbf{a}=-2\mathbf{H}$, $Y_{\mathbf{b}}$ is the gradient-like vector field associated with $\mathbf{b}=-\mathbf{V}=\sum_{j}\,\mathbf{v}_{j}^{\dagger}\,\mathbf{v}_{j}$, and $Z_{\mathcal{K}}$ is a vector field associated with the linear map $\mathcal{K}(\xi)=\sum_{j}\,\mathbf{v}_{j}\,\xi\,\mathbf{v}_{j}^{\dagger}$.

\begin{pf}
We have to prove that $\mathcal{L}_{\widetilde{\Gamma}}\,\mathcal{I}_{\mathfrak{T}_{1}}\subset \mathcal{I}_{\mathfrak{T}_{1}}$, where $\mathcal{I}_{\mathfrak{T}_{1}}\subset\mathcal{F}(\obsp^{*})$ is the ideal of smooth functions vanishing on $\mathfrak{T}_{1}$.

Let us  start recalling equation (\ref{eqn: gradient-like vector fields on the self-adjoint part of the dual of the algebra}), so that we can introduce the gradient-like vector field $\widetilde{\mathbb{Y}}_{\mathbf{b}}$ associated with the symmetric bivector field $\widetilde{\mathcal{R}}$:

\be
\widetilde{\mathbb{Y}}_{\mathbf{b}}=\mathbb{Y}_{\mathbf{b}} -  f_{\mathbf{b}} \,\widetilde{\Delta}\,.
\ee
From this, it follows that $\mathbb{Y}_{\mathbf{b}}= \widetilde{\mathbb{Y}}_{\mathbf{b}} + f_{\mathbf{b}}\,\widetilde{\Delta}$, and thus:

\be\label{eqn: useful decomposition of GKLS on the self-adjoint part of the dual of the algebra}
\widetilde{\Gamma}= \widetilde{\mathbb{X}}_{\mathbf{a}} + \mathbb{Y}_{\mathbf{b}} + \mathbb{Z}_{\mathcal{K}}=\widetilde{\mathbb{X}}_{\mathbf{a}}  + \widetilde{\mathbb{Y}}_{\mathbf{b}} +  f_{\mathbf{b}} \,\widetilde{\Delta} + \mathbb{Z}_{\mathcal{K}}\equiv \widetilde{\mathbb{X}}_{\mathbf{a}}  + \widetilde{\mathbb{Y}}_{\mathbf{b}} +  \widetilde{\mathbb{Z}}_{\mathcal{K}}\,.
\ee
We already know that $\widetilde{\mathbb{X}}_{\mathbf{a}}$ and $\widetilde{\mathbb{Y}}_{\mathbf{b}}$ define, separately, derivations of $\mathcal{F}(\mathfrak{T}_{1})$.
In particular, we know that the derivation associated with $\widetilde{\mathbb{X}}_{\mathbf{a}}$ is the Hamiltonian vector field $X_{\mathbf{a}}$, while the derivation associated with  $\widetilde{\mathbb{Y}}_{\mathbf{b}}$ is the gradient-like vector field $Y_{\mathbf{b}}$.

In order to better understand  $\widetilde{\mathbb{Z}}_{\mathcal{K}}$, we start writing the map $\mathcal{K}(\xi)$ as:

\be
\mathcal{K}(\xi)=Tr\left(\mathcal{K}(\xi)\mathbf{e}^{0}\right)\mathbf{e}_{0} + Tr\left(\mathcal{K}(\xi)\mathbf{e}^{k}\right)\mathbf{e}_{k}\equiv A(\xi) + B(\xi)\,,
\ee
from which it follows that:

\be
\mathbb{Z}_{\mathcal{K}}=\mathbb{Z}_{A} + \mathbb{Z}_{B}\,.
\ee
Next, we look at the map $A$:

\be
A(\xi)=\sum_{j} Tr\left(\mathbf{v}_{j}\,\xi\,\mathbf{v}_{j}^{\dagger}\,\mathbf{e}_{0}\right)\mathbf{e}_{0}=\sum_{j} Tr\left(\mathbf{v}_{j}^{\dagger}\mathbf{v}_{j}\,\xi\right)\frac{\mathbf{e}_{0}}{\sqrt{n}}=\frac{f_{\mathbf{V}}(\xi)}{\sqrt{n}}\,\mathbf{e}_{0}\,.
\ee
Recalling that $\widetilde{\mathbb{Z}}_{\mathcal{K}}=\mathbb{Z}_{\mathcal{K}} - f_{\mathbf{V}}\,\widetilde{\Delta}$, we have:

\be
\widetilde{\mathbb{Z}}_{\mathcal{K}}=\mathbb{Z}_{A} + \mathbb{Z}_{B} - f_{\mathbf{V}}\,\widetilde{\Delta}=f_{\mathbf{V}}\left(\frac{1}{\sqrt{n}}-x^{0}\right)\frac{\partial}{\partial x^{0}} - f_{\mathbf{V}}\,x^{k}\frac{\partial}{\partial x^{k}} + \mathbb{Z}_{B}\,.
\ee
The first term in the RHS clearly vanishes when we are on the hyperplane $x^{0}=\frac{1}{\sqrt{n}}$ representing $\mathfrak{T}_{1}$ in $\obsp^{*}$.
This means that it defines a derivation of $\mathcal{F}(\mathfrak{T}_{1})$ corresponding to the zero vector field.
Furthermore, since the second and third terms in the RHS have no component along $\frac{\partial}{\partial x^{0}}$, they define, separately, derivations of the algebra $\mathcal{F}(\mathfrak{T}_{1})$, that is, vector fields on $\mathfrak{T}_{1}$.
We denote with $Z_{B}$ the vector field on $\mathfrak{T}_{1}$ which is associated with  the vector field  $\mathbb{Z}_{B}$ on $\obsp^{*}$, and with $f_{\mathbf{V}}  \Delta$ the vector field on $\mathfrak{T}_{1}$ which is associated with  the vector field  $f_{\mathbf{V}}\,x^{k}\frac{\partial}{\partial x^{k}}$ on $\obsp^{*}$.
The coordinate expression of $Z_{B}$ reads:

\be
Z_{B}=\mathcal{K}_{\mu}^{k}\,x^{\mu}\, \frac{\partial}{\partial x^{k}}= Tr\left(\mathcal{K}(\mathbf{e}_{\mu})\mathbf{e}^{k}\right)\,x^{\mu}\, \frac{\partial}{\partial x^{k}}\,,
\ee
where $x^{0}=\frac{1}{\sqrt{n}}$ is implicitely assumed.
In the end, $\widetilde{\mathbb{Z}}_{\mathcal{K}}$ defines a derivation of $\mathcal{F}(\mathfrak{T}_{1})$ given by:

\be
Z_{\mathcal{K}}= Z_{B}   -  f_{\mathbf{V}}  \Delta \,.
\ee

Now:

$$
\widetilde{\Gamma}=  \widetilde{\mathbb{X}}_{\mathbf{a}}  + \widetilde{\mathbb{Y}}_{\mathbf{b}} +  \widetilde{\mathbb{Z}}_{\mathcal{K}}
$$
is the sum of three vector fields defining, separately, derivations of $\mathcal{F}(\mathfrak{T}_{1})$, and thus, $\widetilde{\Gamma}$ itself defines a derivation of $\mathcal{F}(\mathfrak{T}_{1})$ which we denote with $\Gamma$.

Eventually, we find that the quantum dynamical evolution generated by the GKLS generator $\mathbf{L}$ of equation (\ref{eqn: K-L equation}) is described by the following GKLS vector field $\Gamma$ on $\mathfrak{T}_{1}$:

\be
\Gamma = X_{\mathbf{a}} + Y_{\mathbf{b}} + Z_{\mathcal{K}}\,.
\ee
\end{pf}
\end{prop}

By construction, the integral curves of  $X_{\mathbf{a}}  + Y_{\mathbf{b}}$ starting at $\rho\in\stsp_{k}\subset\stsp$ remain in $\stsp_{k}$, and thus, it is the vector field $Z_{\mathcal{K}}$ which is responsible for the change of the rank of a quantum state.

Note that $Z_{\mathcal{K}}$, as well as $Y_{\mathbf{b}}$, contains a quadratic term  with respect to the coordinate system $\{x^{k}\}_{k=1,...,n^{2}-1}$ adapted to $\mathfrak{T}_{1}$ given by $f_{\mathbf{V}}\Delta$.
Interestingly, these quadratic terms cancel out in the sum $Y_{\mathbf{b}} + Z_{\mathcal{K}}$, and thus, the GKLS vector field $\Gamma$ representing a linear quantum dynamical process is an affine vector field on $\mathfrak{T}_{1}$.

\vsp

Inspired by the explicit form of $Z_{\mathcal{K}}$, we will give a general prescription to associate a vector field on $\mathfrak{T}_{1}$ with a CPTP map on $\obsp^{*}$.
Let 

$$
A(\xi)=\sum_{j}\,\mathbf{a}_{j}\,\xi\,\mathbf{a}_{j}^{\dagger}\,
$$
be a CPTP map from $\obsp^{*}$ to $\obsp^{*}$.
Next,  define $A^{\sharp}\colon \obsp^{*}\rightarrow\obsp^{*}$ as follows:

\be
A^{\sharp}(\xi)=\sum_{j}\,\mathbf{a}_{j}^{\dagger}\,\xi\,\mathbf{a}_{j}\,.
\ee
It is clear that $A^{\sharp}$ is  a completely-positive map according to Choi's theorem  \cite{choi-completely_positive_linear_maps_on_complex_matrices}.
Now, we set:

\be
Z_{A}:=\left(A_{\mu}^{k}\,x^{^{\mu}}  - x^{k} \,f_{A^{\sharp}(\mathbf{e}_{0})}\right) \frac{\partial}{\partial x^{k}}\,,
\ee
where $x^{0}=\frac{1}{\sqrt{n}}$ is implicitely assumed.

Note that this way of associating a vector field $Z_{A}$ on $\mathfrak{T}_{1}$ with a CPTP map $A$ on $\obsp^{*}$ is completely unrelated with $A$ being the CPTP map of some GKLS generator.

\vsp

As said before, both $Y_{\mathbf{b}}$ and $Z_{A}$ contain, in general, non-affine parts with respect to the coordinate system $\{x^{k}\}_{k=1,...,n^{2}-1}$ adapted to $\mathfrak{T}_{1}$.
This means that their sum $Y_{\mathbf{b}} + Z_{A}$ is, in general, a non-affine vector field on $\mathfrak{T}_{1}$.
However,  if we take 

\be\label{eqn: fine-tuning 1}
\mathbf{b}=-\sum_{j=1}^{N}\,\mathbf{v}^{\dagger}_{j}\,\mathbf{v}_{j}\,,
\ee
and

\be\label{eqn: fine-tuning 2}
A(\xi)=\sum_{j=1}^{N}\,\mathbf{v}_{j}\,\xi\,\mathbf{v}_{j}^{\dagger}\,,
\ee
then the non-affine terms in $Y_{\mathbf{b}}$ and $Z_{A}$ cancel each other, and $Y_{\mathbf{b}} + Z_{A}$ becomes an affine vector field.
This is precisely what happened in the construction of the GKLS vector field $\Gamma$ of equation (\ref{eqn: GKLS vector field}).
We can not describe a linear quantum dynamical evolution using the vector field $\Gamma=X_{\mathbf{a}} + Y_{\mathbf{b}} + Z_{A}$, where $\mathbf{a},\mathbf{b}$ and $A$ are completely arbitrary.
The linearity requirement for the evolution, which is equivalent to $\Gamma$ being an affine vector field on $\mathfrak{T}_{1}$, forces us to  fine-tune $Y_{\mathbf{b}}$ and $Z_{A}$ using equations (\ref{eqn: fine-tuning 1}) and (\ref{eqn: fine-tuning 2}).

\vsp

\begin{exmp}[Phase damping of a qubit]\label{example: Phase damping of a qubit}

We will now give the explicit expression of the GKLS vector field associated with the quantum dynamical process known as the phase damping of a qubit.
For the notation, we refer to example  \ref{example: two level quantum system geometry}.

The GKLS generator for the phase damping is given by equation (\ref{eqn: K-L equation}) with $\mathbf{H}=0$, $N=1$,  and $\mathbf{v}\equiv\mathbf{v}_{1}=\sqrt{\gamma}\,\sigma_{3}$:

\be
\mathbf{L}(\rho)=- \gamma \left(\rho - \sigma_{3}\,\rho\,\sigma_{3}\right)\,.
\ee
To compute the GKLS vector field $\Gamma$, note that $\mathbf{a}=2\mathbf{H}=0$ implies $X_{\mathbf{a}}=0$, and, since $-\mathbf{b}=\mathbf{V}=\mathbf{v}^{\dagger}\,\mathbf{v}=\gamma\sigma_{0}$, it is $Y_{\mathbf{b}}=0$.
It follows that $\Gamma = Z_{\mathcal{K}}$.

Now:

\be
Tr\left(\mathcal{K}(\sigma_{\mu})\sigma^{k}\right)= \gamma Tr\left(\sigma_{3}\rho\,\sigma_{3}\,\mathbf{e}^{k}\right)= 2\gamma\,x^{3}\,\delta_{3}^{k} - \gamma \, x^{l}\,\delta_{l}^{k}\,,
\ee
and $f_{\mathbf{V}}= \gamma$, which means: 
:

\be\label{eqn: GKLS vector field phase damping qubit}
\Gamma = Z_{\mathcal{K}} = -2\gamma\left(x^{1}\,\frac{\partial}{\partial x^{1}} + x^{2}\,\frac{\partial}{\partial x^{2}}\right)\,.
\ee
The flow $\Phi_{\tau}$ generated by $\Gamma$ reads:

\be
\Phi_{\tau}(\rho)=\frac{1}{2}\,\left(\sigma_{0} + \exp(-2\gamma\tau)\left(x^{1}\sigma_{1} + x^{2}\sigma_{2}\right) + x^{3}\sigma_{3}\right)\,,
\ee
and it is clear that this dynamics only affects the phase terms (off-diagonal terms) of $\rho$ represented by its components along $\sigma_{1}$ and $\sigma_{2}$.
All the quantum states lying on the $x^{3}$ axis are fixed points of the dynamics, and it is clear that an initial state $\rho$ will evolve towards its projection on the $x^{3}$ axis.
Indeed, from the geometrical point of view, the integral curves of $\Gamma$ are radial lines in a two-dimensional plane orthogonal to the $x^{3}$-axis.
Therefore, the dynamical evolution of the initial state $\rho$ is always transversal to the spheres centered in $x^{1}=x^{2}=x^{3}=0$.
These spheres represent isospectral  quantum states, hence, the dynamics will change the spectrum of $\rho$ giving rise to dissipation.

Note that, for $\tau\leq0$, the flow of $\Gamma$ takes a state $\rho$ out of $\stsp$, and thus, from the point of view of the space of states $\stsp$, $\Gamma$ generates a one-parameter semigroup of transformations.

\end{exmp}

\begin{exmp}[Energy damping of a qubit]

Let us now look at the dynamical evolution of a qubit associated with a GKLS generator having $\mathbf{H}=0$, $N=1$, and $\mathbf{v}\equiv\mathbf{v}_{1}=\sqrt{\gamma}\left(\sigma_{1} + \imath\sigma_{2}\right)$:

\be
\mathbf{L}(\rho)=-\mathbf{V}\odot\rho + \gamma\left(\sigma_{1} + \imath\sigma_{2}\right)\,\rho\,\left(\sigma - \imath\sigma_{2}\right)\,.
\ee
Now, $\mathbf{a}=2\mathbf{H}=0\Rightarrow X_{\mathbf{a}}=0$ as for the phase damping, however, $\mathbf{b}=-\mathbf{V}=\mathbf{v}^{\dagger}\mathbf{v}=-2\gamma(\sigma_{0} - \sigma_{3})$, and thus the gradient-like vector field $Y_{\mathbf{b}}$ reads:

\be
Y_{\mathbf{b}}=  2\gamma\,\left(\frac{\partial}{\partial x^{3}} -  x^{3}\,\Delta\right)\,,
\ee
with $\Delta$ the dilation vector field, and we see that it has a quadratic term.
As stated before, we will see that the vector field $Z_{\mathcal{K}}$ contains a quadratic term which will cancel the quadratic term of $Y_{\mathbf{b}}$.
This is a concrete instance of the fine-tuning between $Y_{\mathbf{b}}$ and $Z_{\mathcal{K}}$ imposed by the requirement of linearity for the quantum dynamics.

Now, we have:

$$
Tr\left(\mathbf{v}\,\rho\,\mathbf{v}^{\dagger}\,\sigma^{k}\right)=\frac{1}{2}\left(Tr\left(\mathbf{v}\,\mathbf{v}^{\dagger}\,\sigma^{k}\right) + x^{l}\,Tr\left(\mathbf{v}\,\sigma_{l}\,\mathbf{v}^{\dagger}\,\sigma^{k}\right)\right)=
$$
\be
=2\gamma\left( x^{l}(\delta_{1l}\delta^{k}_{1} + \delta_{2l}\delta^{k}_{2} - \delta_{l}^{k}) + \delta^{k}_{3}\right)\,
\ee
and $f_{\mathbf{V}}=2\gamma\left(1 - x^{3}\right)$. 
Therefore:

\be
Z_{\mathcal{K}}=2\gamma\left(1- x^{3} \right)\,\frac{\partial}{\partial x^{3}} - 2\gamma\,\left(1 - x^{3}\right)\Delta \,.
\ee
Collecting the results we obtain the following form for the GKLS vector field

\be
\Gamma = Z_{\mathcal{K}} + Y_{\mathbf{b}}=-2\gamma\left(x^{1}\,\frac{\partial}{\partial x^{1}} + x^{2}\,\frac{\partial}{\partial x^{2}}\right) + 4\gamma\left(1 - x^{3}\right)\,\frac{\partial}{\partial x^{3}}\,.
\ee
The quadratic terms in $Z_{\mathcal{K}}$ and $Y_{\mathbf{b}}$ canceled out, and we are left with  an affine vector field as it should be.
We stress the fact that this cancellation occurs because of the fine-tuning between $Z_{\mathcal{K}}$ and $Y_{\mathbf{b}}$ imposed by the linearity requirement for the quantum evolution.

Looking at $\Gamma$, we immediately see that it has a single fixed point, specifically, the pure state $\rho=\frac{1}{2}(\sigma_{0} + \sigma_{3})$.
Furthermore, we realize that $\Gamma$ is the sum of the GKLS vector field of the phase damping with the vector field $4\gamma\left(1 - x^{3}\right)\,\frac{\partial}{\partial x^{3}}$.
These two vector fields commute, and thus the flow $\Phi_{\tau}$ of their sum can be written as the composition of their flows.
The specific expression is:

\be
\Phi_{\tau}(\rho)=\frac{1}{2}\,\left(\sigma_{0} + \mathrm{e}^{-\gamma\tau}\left(x^{1}\sigma_{1} + x^{2}\sigma_{2}\right) + \left(\mathrm{e}^{-4\gamma\tau}\left( x^{3} - 1\right) + 1\right)\sigma_{3}\right)\,.
\ee
The asymptotic behaviour of this dynamics is quite interesting.
Indeed, every initial state $\rho$ evolves toward a common asymptotic state $\rho_{\infty}$, specifically, the pure state $\rho_{\infty}=\frac{1}{2}(\sigma_{0} + \sigma_{3})$.
On the one hand, if we start from an initial state $\rho$ which is mixed, the dynamical evolution may be read as a ``purification'' process for $\rho$, which, however, is accomplished only in the limit $\tau\rightarrow + \infty$.
On the other hand, if we start from an initial state $\rho$ which is pure, the dynamics will immediately destroy its purity turning it into a mixed state, and then it will start to ``purify'' it again.
We see here a ``collapse'' and ``revival'' phenomenon.
\end{exmp}

\section{Quantum random unitary semigroups}\label{sec: Quantum Poisson semigroups}

Now that we have written the GKLS generator of a linear quantum dynamical process in terms of a vector field $\Gamma$ on the manifold $\mathfrak{T}_{1}$, we are able to use all the tools from the theory of dynamical systems in the quantum context.
This will help us, for example, in analyzing the stability properties of a given quantum dynamical process.

First of all, let us recall that the fixed points of the dynamical evolution associated with the GKLS vector field $\Gamma$ are all those points $\xi_{f}$ such that $\Gamma(\xi_{f})=0$.
Denoting with $\Phi_{\tau}$ the flow of $\Gamma$, we have that $\Phi_{\tau}(\xi_{f})=\xi_{f}$ for all $\tau$.
Fixed points are of primary interest in the stability theory of fixed points of dynamical systems.
Essentially, given a fixed point $\xi_{f}$, stability theory consists of understanding the long time behaviour of the dynamical trajectories with initial conditions belonging to a neighbourhood of $\xi_{f}$.

The literature on the subject is mainly focused on classical physical systems.
What is interesting, is that the geometric reformulation of quantum dynamics we have achieved in the previous section allows us to make good use of the results of stability theory directly in the quantum case.

We do not want to enter into a detailed and exhaustive discussion of the stability theory of quantum dynamical evolutions.
Our main scope is to show how the geometric formulation of the GKLS dynamics can be used to gain physical intuition on  quantum situations using mathematical tools coming from classical physics.

\vsp

Let us start recalling  the so-called LaSalle invariance principle \cite{lasalle-some_extensions_of_liapunov_second_method,abraham_marsden_ratiu-manifolds_tensor_analysis_and_applications}:

\begin{thm}\label{thm: lasalle invariance principle}
Let $B$ be be a finite-dimensional Banach space, and let $M\subseteq B$ be a finite-dimensional differential manifold. 
Consider the smooth dynamical system associated with the complete vector field $\Gamma$.
Let $\Omega$ be a compact set in $M$ that is invariant under the flow $\Phi_{\tau}$ of $\Gamma$ for $\tau\geq0$.
Let $F\colon M\rightarrow\mathbb{R}$ be a smooth function such that $F\geq0$ on $\Omega$, and assume that:

\be
\mathcal{L}_{\Gamma}F\leq 0
\ee
on $\Omega$.
Let $S_{\infty}$ be the largest invariant set in $\Omega$, for $\tau\in\mathbb{R}$, where $\mathcal{L}_{\Gamma}F=0$.
If $x\in\Omega$, then:

\be
\lim_{\tau\rightarrow+\infty}\,\left(\inf_{m^{*}\in S_{\infty}}\,\left|\left|\Phi_{\tau}(m) - m^{*}\right|\right|\right)=0
\ee 
where $||\cdot||$ is the norm of $B$ .
In particular, if $S_{\infty}$ is an isolated fixed point of $\Gamma$, it is asymptotically stable.
\end{thm}

We call a function $F$ satisfying the hypotesis of theorem \ref{thm: lasalle invariance principle} a LaSalle function for the vector field $\Gamma$.

Referring to theorem \ref{thm: lasalle invariance principle}, we take $B=\obsp^{*}$, $M=\mathfrak{T}_{1}$, $\Gamma$ the GKLS vector field of the semigroup at hand, and $\Omega=\stsp$, where $\stsp$ is the space of quantum states.
We will denote with $\rho$ a generic element in $\stsp$.
Consider the function:

\be
F(\xi):=\frac{\chi(\xi)}{2}=\frac{Tr(\xi^{2})}{2}\,,
\ee
where $\chi(\xi)=Tr(\xi^{2})$ is the  purity function.
This is a smooth function on $\mathfrak{T}_{1}$ such that $F(\rho)\geq0$.
It is connected to the so-called linearized entropy function:

\be
S_{L}(\xi)=1-\chi(\xi)=1-2F(\xi)\,.
\ee
We will analyze the so-called quantum Poisson semigroups, quantum Gaussian semigroups, and quantum random unitary semigroups \cite{lindblad-on_the_generators_of_quantum_dynamical_semigroups, kossakowski-on_quantum_statistical_mechanics_of_non_hamiltonian_systems, aniello_kossakowski_marmo_ventriglia-quantum_brownian_motion_on_lie_groups_and_open_quantum_systems}, and we will show that the purity function $\chi(\rho)=Tr(\rho^{2})$ is a LaSalle function for the GKLS vector field associated with these semigroups in every dimension.

The GKLS generator $\mathbf{L}$ for the quantum random unitary semigroups is characterized by the following form \cite{aniello_kossakowski_marmo_ventriglia-quantum_brownian_motion_on_lie_groups_and_open_quantum_systems} :

\be
\mathbf{L}(\xi)=-2[[\mathbf{H}\,,\xi]] - \mathbf{V}\odot\xi + \sum_{j=1}^{n^{2}-1}\,\alpha_{j}\mathbf{e}_{j}\,\xi\,\mathbf{e}_{j} + \beta\sum_{j=1}^{r\leq n^{2}-1}\,p_{j}\mathbf{U}_{j}\,\xi\,\mathbf{U}_{j}^{\dagger}\,,
\ee
where $\mathbf{H}\in\mathcal{B}(\mathcal{H})$ is self-adjoint, $\alpha,\beta$ are non-negative real numbers, $\{p_{j}\}_{j=1,..r}$ is a probability vector, $\{\mathbf{e}_{j}\}_{j=1,...,n^{2}-1}$ is an orthonormal set of self-adjoint operators in $\mathcal{B}(\mathcal{H})$, $\mathbf{V}=\sum_{j=1}^{n^{2}-1}\mathbf{e}_{j}^{2}$, and $U_{j}$ is unitary for all $j$.

We will break the problem in steps.
We will start analyzing the so-called quantum Poisson semigroups.
These form a subclass of the quantum random unitary semigroups for which $N=1$, and $\mathbf{v}\equiv\mathbf{v}_{1}=\mathbf{U}$ is unitary:

\be
\mathbf{L}(\xi)= - \xi +  \mathbf{U}\,\xi\, \mathbf{U}^{\dagger}\,.
\ee
Next, we will consider ``positive linear combinations'' of quantum Poisson semigroups, and add to them a Hamiltonian term.

After the quantum Poisson semigroups, we will focus on the so-called quantum Gaussian semigroups.
Again, these semigroups form a subclass of the quantum random unitary semigroups.
Their GKLS generator is characterized by $N=1$ and $\mathbf{v}\equiv \mathbf{v}_{1}$ self-adjoint:

\be
\mathbf{L}(\xi)=-\mathbf{v}^{2}\odot\xi + \mathbf{v}\,\xi\,\mathbf{v}\,.
\ee
We will proceed considering ``positive linear combinations'' of quantum Gaussian semigroups, and adding to them a Hamiltonian term.

Finally, we will consider the general case of quantum random unitary semigroups.

\subsection*{Quantum Poisson semigroups}

Quantum Poisson semigroups are characterized by a GKLS generator $\mathbf{L}$ with $\mathbf{H}=0$, and with a single, unitary Kraus operator $\mathbf{v}=\mathbf{U}$ \cite{lindblad-on_the_generators_of_quantum_dynamical_semigroups, kossakowski-on_quantum_statistical_mechanics_of_non_hamiltonian_systems, aniello_kossakowski_marmo_ventriglia-quantum_brownian_motion_on_lie_groups_and_open_quantum_systems}:

\be
\mathbf{L}(\xi)= - \xi +  \mathbf{U}\,\xi\, \mathbf{U}^{\dagger}\,.
\ee
Since $\mathbf{V}=\mathbf{U}^{\dagger}\,\mathbf{U}=\mathbb{I}$, it follows from equation (\ref{eqn: gradient-like vector field on trace 1}) that the gradient-like vector field $Y_{\mathbf{b}}$ in the GKLS vector field $\Gamma$ describing $\mathbf{L}$ is zero.
Being $\mathbf{H}=\mathbf{0}$, the Hamiltonian vector field $X_{\mathbf{a}}$ is zero too, and we are left only with the vector field $Z_{\mathcal{K}}$.
Concerning this vector field, we note that $f_{\mathbf{V}}=1$, so that:

\be
Z_{K}=Z_{B}  -   \Delta\,.
\ee
Then, we note that:

\be
B(\xi)=Tr\left(\mathbf{U}\,\xi\,\mathbf{U}^{\dagger}\,\mathbf{e}^{k}\right)\mathbf{e}_{k}\, \,,
\ee
and thus:

\be\label{eqn: GKLS vector field for Poisson semigroups}
\Gamma = Z_{K} = Z_{B}  -  \Delta = Tr\left(\mathbf{U}\,\mathbf{e}_{l}\,\mathbf{U}^{\dagger}\,\mathbf{e}^{k}\right)x^{l}\,\frac{\partial}{\partial x^{k}} -  \Delta \,,
\ee
where $\xi=\frac{\mathbf{e}_{0}}{\sqrt{n}} + x^{l}\mathbf{e}_{l}$.
An easy calculation, shows that the fixed points  for the dynamical system associated with $\Gamma$ are all those $\xi_{f}$ such that $[\mathbf{U}\,,\xi_{f}]=0$, in particular,  the maximally mixed state $\rho=\frac{\mathbf{e}_{0}}{\sqrt{n}}=\frac{\mathbb{I}}{n}$ is a fixed point for every choiche of the operator $\mathbf{U}$.

We will show that $F(\xi)$ is a LaSalle function for $\Gamma$.

\begin{prop}
The function:

\be
F(\xi)=\frac{Tr\left(\xi^{2}\right)}{2}
\ee
is a LaSalle function for the GKLS vector field $\Gamma$ of equation (\ref{eqn: GKLS vector field for Poisson semigroups}) representing a quantum Poisson semigroup.

\begin{pf}
Writing 

\be
\rho= \frac{1}{\sqrt{n}}\mathbf{e}_{0} + \tilde{\rho}=\frac{1}{\sqrt{n}}\mathbf{e}_{0} + x^{j}\mathbf{e}_{j}\,,
\ee
it is:

\be
F(\rho)=\frac{1}{2n} + \frac{\delta_{jk}x^{j}x^{k}}{2}\,.
\ee
Therefore:

\be\label{eqn: derivative of purity}
\left.\frac{\partial F}{\partial x^{k}}\right|_{\rho}=\delta_{jk}x^{j}=Tr(\rho\,\mathbf{e}_{k})\,.
\ee
Consequently:

\be
\left.\mathcal{L}_{\Gamma}F\right|_{\rho}=\frac{\partial F}{\partial x^{k}}\,\Gamma^{k}=\delta_{jk}x^{j}\left(Tr\left(\mathbf{U}\,\mathbf{e}_{l}\,\mathbf{U}^{\dagger}\,\mathbf{e}^{k}\right)x^{l} - x^{k}\right)=Tr\left(\mathbf{U}\,\tilde{\rho}\,\mathbf{U}^{\dagger}\,\tilde{\rho}\right) - Tr\left(\tilde{\rho}\,\tilde{\rho}\right)\,.
\ee
By direct computation, we see that:

\be
Tr\left(\mathbf{U}\,\rho\,\mathbf{U}^{\dagger}\,\rho\right) - Tr\left(\rho\,\rho\right)= Tr\left(\mathbf{U}\,\tilde{\rho}\,\mathbf{U}^{\dagger}\,\tilde{\rho}\right) - Tr\left(\tilde{\rho}\,\tilde{\rho}\right)\,.
\ee
Writing $\rho_{U}=\mathbf{U}\,\rho\,\mathbf{U}^{\dagger}$, we have:

\be
\left.\mathcal{L}_{\Gamma}F\right|_{\rho}=Tr\left(\rho_{U}\,\rho\right) - Tr\left(\rho\,\rho\right)\,.
\ee
The expression $Tr\left(\rho_{U}\,\rho\right)$ on the RHS is nothing but the Euclidean scalar product,  in the Euclidean vector space $\obsp^{*}$, between the vectors $\rho_{U}$ and $\rho$.
Analogously, $Tr\left(\rho\,\rho\right)\equiv|\rho|^{2}$ is the scalar product between $\rho$ and itself.
Therefore:

\be
\left.\mathcal{L}_{\Gamma}F\right|_{\rho}=|\rho_{U}|\,|\rho|\,\cos(\theta) - |\rho|^{2}\,,
\ee
where $\theta$ is the angle between $\rho$ and $\rho_{U}$.
Being $\rho_{U}=\mathbf{U}\,\rho\,\mathbf{U}^{\dagger}$, it follows that $|\rho_{U}|=|\rho|$ and thus:

\be\label{eqn: derivative of the lasalle function for poisson semigroups}
\left.\mathcal{L}_{\Gamma}F\right|_{\rho}=|\rho|^{2}\,\left(\cos(\theta) - 1\right)\leq 0\,,
\ee
where the equality holds if and only if $\rho_{U}=\rho$.
This means that theorem \ref{thm: lasalle invariance principle} applies, and thus $F$ is a LaSalle function for $\Gamma$ as claimed.
\end{pf}
\end{prop}

According to theorem \ref{thm: lasalle invariance principle}, the accumulating set $S_{\infty}$ is the largest invariant subset in:

\be
E:=\left\{\rho\in\stsp\colon \left.\mathcal{L}_{\Gamma}F\right|_{\rho}=0\right\}\,.
\ee
From equation (\ref{eqn: derivative of the lasalle function for poisson semigroups}), 
we have that $E$ coincide with the intersection of the space of states $\stsp$ with the set of fixed points of the GKLS vector field $\Gamma$.
Since every $\rho\in E$ is a fixed point, the set $E$ is an invariant set for the dynamics, and thus $E=S_{\infty}$.
From the practical point of view, we can see that $S_{\infty}$   is the intersection of the commutant\footnote{The commutant $\mathcal{C}_{\mathbf{A}}$ of $\mathbf{A}\in\mathcal{B}(\mathcal{H})$ is the set of all elements in $\mathcal{B}(\mathcal{H})$ commuting with $\mathbf{A}$.} $\mathcal{C}_{\mathbf{U}}$ of $\mathbf{U}$ with the space of states $\stsp$.
The comutant $\mathcal{C}_{\mathbf{U}}$ is a vector space (actually, an algebra with respect to the operator product) the dimension $d_{U}$ of which depends on the degenerancy of the spectrum of $\mathbf{U}$.
Specifically, it is:

\be
d_{U}=\sum_{j=1}^{m}(d_{j})^{2}\,,
\ee
where $m$ is the number of different eigenvalues of $\mathbf{U}$, and $d_{j}$ denotes the  degenerancy of the $j$-th eigenvalue of $\mathbf{U}$.
Consequently, the more degenerancy in the spectrum of $\mathbf{U}$, the bigger is the accumulating set $S_{\infty}$.

\begin{exmp}[Phase damping of a qubit revisited I]\label{example: Phase damping of a qubit revisited I}
Let us come back to the phase damping studied in example \ref{example: Phase damping of a qubit}.
Because of the peculiar properties of the Pauli matrices, we have that $\mathbf{v}$ is unitary when we set $\gamma=1$.
The GKLS vector field in equation (\ref{eqn: GKLS vector field phase damping qubit}) becomes:

\be
\Gamma = -2\left(x^{1}\,\frac{\partial}{\partial x^{1}} + x^{2}\,\frac{\partial}{\partial x^{2}}\right)\,, 
\ee
and its associated flow $\Phi_{\tau}$ is:

\be
\Phi_{\tau}(\rho)=\frac{1}{2}\,\left(\sigma_{0} + \exp(- 2\tau)\left(x^{1}\sigma_{1} + x^{2}\sigma_{2}\right) + x^{3}\sigma_{3}\right)\,.
\ee
It is easy to see that the set $S_{\infty}$ is precisely the intersection between the $x^{3}$-axis and the Bloch-ball.
Furthermore, from the explicit form of $\Phi_{\tau}$ it follows that the initial state $\rho$ tends to the state $\rho_{\infty}$ which is the projection of $\rho$ onto the $x^{3}$-axis.
\end{exmp}

We can go a little further and analyze the ``positive linear combinations'' of quantum Poisson semigroups with a Hamiltonian term.
These are all those quantum dynamics characterized by a GKLS generator $\mathbf{L}$ for which $\mathbf{H}\neq\mathbf{0}$, $\mathbf{v}_{j}=\alpha_{j}\mathbf{U}_{j}$ with $\mathbf{U}_{j}$ unitary and $\alpha_{j}\in\mathbb{C}$ for all $j$.

\be\label{eqn: GKLS generator for linear combination of quantum poisson semigroups with hamiltonian}
\mathbf{L}(\xi)= -2\left[\left[\mathbf{H}\,,\xi\right]\right] -  \sum_{j=1}^{N}\,|\alpha_{j}|^{2}\xi +  \sum_{j=1}^{N}\,|\alpha_{j}|^{2}\mathbf{U}_{j}\,\xi\, \mathbf{U}^{\dagger}_{j}\,.
\ee
Again, there is no gradient-like contribution in the GKLS vector field $\Gamma$, however, being $\mathbf{H}\neq\mathbf{0}$, there is a Hamiltonian contribution.

An explicit calculation shows that:

$$
\Gamma = \sum_{j} |\alpha_{j}|^{2}\, \Gamma_{j} -2X_{\mathbf{H}} =
$$
\be\label{eqn: GKLS vector field for linear combinations of quantum poisson with hamiltonian}
=\sum_{j} |\alpha_{j}|^{2}\left(Tr\left(\mathbf{U}_{j}\,\mathbf{e}_{l}\,\mathbf{U}^{\dagger}_{j}\,\mathbf{e}^{k}\right)x^{l}\,\frac{\partial}{\partial x^{k}} -  \Delta\right) -
 2Tr\left([[\mathbf{H}\,,\mathbf{e}_{l}]]\,\mathbf{e}^{k}\right)x^{l}\,\frac{\partial}{\partial x^{k}}  \,.
\ee
where $\Gamma_{j}$ is the GKLS vector field of the quantum Poisson semigroup associated with $\mathbf{U}_{j}$.
The fixed points of $\Gamma$ are now all those $\xi$ such that:

\be
\sum_{j} |\alpha_{j}|^{2}\left(\mathbf{U}\,\xi\,\mathbf{U}^{\dagger} -\xi\right)= 2[[\mathbf{H}\,,\xi]]\,.
\ee

\begin{prop}\label{prop: purity is a lasalle function for linear combinations of quantum poisson semigroups with hamiltonian}
The function:

\be
F(\xi)=\frac{Tr\left(\xi^{2}\right)}{2}
\ee
is a LaSalle function for the GKLS vector field $\Gamma$ of equation (\ref{eqn: GKLS vector field for linear combinations of quantum poisson with hamiltonian}).

\begin{pf}
Because of the linearity of the Lie derivative, we have:
\be
\left.\mathcal{L}_{\Gamma}F\right|_{\rho}=\sum_{j} |\alpha_{j}|^{2}\, \left( \left.\mathcal{L}_{\Gamma_{j}}F\right|_{\rho}\right) -2 \left.\mathcal{L}_{X_{\mathbf{H}}}F\right|_{\rho}\,.
\ee
The Lie derivative of the function $F(\xi)=\frac{Tr(\xi\,\xi)}{2}$ with respect to $X_{\mathbf{H}}$ is easily seen to be zero.
To see this, recall that the flow of the Hamiltonian vector field $X_{\mathbf{H}}$ is given by 

\be
\Phi_{\tau}(\xi)=\exp(-\imath \tau\mathbf{H})\,\xi\,\exp(\imath \tau \mathbf{H})\,,
\ee
and thus $\xi$ and $\Phi_{\tau}(\xi)$ have the same eigenvalues for all $\tau$.
Now, we can write 

$$
F(\xi)=\sum_{k}\lambda^{2}_{k}\,,
$$
with $\lambda_{k}$ the $k$-th eigenvalue of $\xi$, from which it follows that 

$$
F(\xi)=F(\Phi_{\tau}(\xi))
$$
for all $\tau$, which is equivalent to $\mathcal{L}_{X_{\mathbf{H}}}F=0$.
Consequently:

\be\label{eq: lie derivative of the purity with respect to the dynamical vector field of the sum of Poisson semigroups}
\left.\mathcal{L}_{\Gamma}F\right|_{\rho}=\sum_{j} |\alpha_{j}|^{2}\, \left( \left.\mathcal{L}_{\Gamma_{j}}F\right|_{\rho}\right) -2 \left.\mathcal{L}_{X_{\mathbf{H}}}F\right|_{\rho}=\sum_{j}\,|\alpha_{j}|^{2}\,|\rho|^{2}\,\left(\cos(\theta_{j}) - 1\right)\leq 0\,,
\ee
where $\theta_{j}$ is the angle between $\rho$ and $\rho_{U_{j}}=\mathbf{U}_{j}\,\rho\,\mathbf{U}_{j}^{\dagger}$.
This means that theorem \ref{thm: lasalle invariance principle} applies, and the proposition is proved.
\end{pf}
\end{prop}

Note that the Lie derivative in equation (\ref{eq: lie derivative of the purity with respect to the dynamical vector field of the sum of Poisson semigroups}) is  zero if and only if $\rho$  commutes with $\mathbf{U}_{j}$ for all $j$, and thus:

\be
E:=\left\{\rho\in\stsp\colon \left.\mathcal{L}_{\Gamma}F\right|_{\rho}=0\right\}\,,
\ee
coincides with the intersection of $\stsp$ with the intersection of the commutants $\mathcal{C}_{\mathbf{U}_{j}}$.
The accumulating set $S_{\infty}$ is then the largest invariant set in $E$.
Clearly, $S_{\infty}$ highly depends on the spectral properties of the unitary operators $\mathbf{U}_{j}$.
One could be tempted to say that the Hamiltonian part plays no role in this discussion since the Lie derivative of the LaSalle function with respect to the Hamiltonian vector field vanishes.
However, we know that $S_{\infty}\subseteq E$ must be an invariant set with respect to the total dynamics of the system, and the Hamiltonian part of $\Gamma$ obviously takes part in determing the explicit form of the dynamical trajectories.
The maximally mixed state is always in $S_{\infty}$.

\begin{rem}
In the qubit case, it is enough to take $N=2$ with $\mathbf{v}_{1},\mathbf{v}_{2}$ any couple of different Pauli matrices (except the identity) in order for $S_{\infty}$ to coincide with the singleton represented by the maximally mixed state.
\end{rem}

\begin{exmp}[Phase damping of a qubit, revisited II]
Let us come back to the phase damping studied in example \ref{example: Phase damping of a qubit revisited I}, and denote with $\Gamma_{U}$ its GKLS vector field.
We want to understand what happens when we add a Hamiltonian term to the GKLS vector field $\Gamma_{U}$.
The resulting GKLS vector field is:

\be
\Gamma = X_{\mathbf{a}} + \Gamma_{U}\,.
\ee
Now, let us write the most general expression for the Hamiltonian vector field $X_{\mathbf{a}}$:

\be
X_{\mathbf{a}}=\left(h_{3} x^{2} - h_{2}x^{3} \right)\frac{\partial}{\partial x^{1}} + \left(h_{3} x^{1} - h_{1}x^{3}\right)\frac{\partial}{\partial x^{2}} + \left(h_{1} x^{2} - h_{2}x^{1}\right)\frac{\partial}{\partial x^{3}}\,,
\ee
where $\mathbf{a}=-2\mathbf{H}=-2h_{\mu}\,\sigma^{\mu}$.
Recall that the fixed points of $\Gamma_{U}$ are all those points commuting with $\mathbf{U}$, while the fixed points of $\mathbf{H}$ are all those points commuting with $\mathbf{H}$.
A direct computation shows that:

\be
[X_{\mathbf{a}}\,,\Gamma_{U}]=h^{2}x^{3}\,\frac{\partial}{\partial x^{1}} + h^{1}x^{3}\,\frac{\partial}{\partial x^{2}}\,.
\ee
From this, it follows that $[X_{\mathbf{a}}\,,\Gamma_{U}]=0$ if and only if $\mathbf{H}=h_{3}\sigma^{3}$.
In this case, all the points in $E$ are fixed points of $\Gamma$ because $[\mathbf{U}\,\mathbf{H}]=\mathbf{0}$.
Consequently, $E=S_{\infty}$ exactly as in example \ref{example: Phase damping of a qubit revisited I}.

On the other hand, if $\mathbf{H}=h_{1}\sigma^{1} + h_{2}\sigma^{2}$, the situation changes drastically.
Indeed, let us take $h_{1}=1$, $h_{2}=h_{3}=0$.
The GKLS vector field becomes:

\be
\Gamma = -2x^{1}\,\frac{\partial}{\partial x^{1}}   -2\left(x^{2} + x^{3}\right)\,\frac{\partial}{\partial x^{2}} + 2x^{2}\,\frac{\partial}{\partial x^{3}}\,.
\ee
This vector field has a single fixed point, namely, the maximally mixed state $\rho_{m}=\frac{\mathbb{I}}{2}$.
Writing $\xi\in\mathfrak{T}_{1}$ as:

$$
\xi=\frac{1}{2}\left(\sigma_{0} + x^{1}\sigma_{1} + x^{2}\sigma_{2} + x^{3}\sigma_{3}\right)\,,
$$
the explicit form of $\Phi_{\tau}$ reads:

\be
\Phi_{\tau}(\xi)=\left\{\begin{array}{ccl} 
x^{1}(\tau) & = & x^{1}\,\mathrm{e}^{-2\tau} \\
 & & \\ 
x^{2}(\tau) & = & -\mathrm{e}^{-\tau}\left(x^{2}\left(\frac{\sin(\sqrt{3}\tau)}{\sqrt{3}} - \cos(\sqrt{3}\tau)\right) + \frac{2 x^{3}}{\sqrt{3}}\,\sin(\sqrt{3}\tau)\right) \\
 & & \\
x^{3}(\tau) & = & \mathrm{e}^{-\tau}\left(\frac{2 x^{2}}{\sqrt{3}}\,\sin(\sqrt{3}\tau) + x^{3}\left(\frac{\sin(\sqrt{3}\tau)}{\sqrt{3}} + \cos(\sqrt{3}\tau)\right)\right) 
\end{array}\right.\,.
\ee

Now, the set $E$ consits of all those quantum states $\rho$ commuting with $\mathbf{U}$.
It is clear that, given $\rho\in E$, it is $\Phi_{\tau}(\rho)\in \rho$ if and only if $\rho=\rho_{m}$ wiht $\rho_{m}=\frac{1}{2}\sigma_{0}$ the maximally mixed state.
Otherwise, $\rho$ is mapped outside $E$ by the dynamical evolution $\Phi_{\tau}$.
This means that the largest invariant set in $E$ is the singleton $\{\rho_{m}\}$, and we conclude that $S_{\infty}=\{\rho_{m}\}$.
The Hamiltonian term $X_{\mathbf{a}}$ has thus changed the long-time behaviour of the dynamics making the maximally mixed state the limiting point of the dynamical evolution of every quantum state $\rho$.
\end{exmp}

\subsection{Quantum Gaussian semigroups}

A quantum Gaussian semigroup \cite{lindblad-on_the_generators_of_quantum_dynamical_semigroups,kossakowski-on_quantum_statistical_mechanics_of_non_hamiltonian_systems,aniello_kossakowski_marmo_ventriglia-quantum_brownian_motion_on_lie_groups_and_open_quantum_systems}
 is characterized by GKLS generator $\mathbf{L}$ having $\mathbf{H}=\mathbf{0}$, $N=1$ and $\mathbf{v}\equiv \mathbf{v}_{1}$ self-adjoint:

\be
\mathbf{L}(\xi)=-\mathbf{v}^{2}\odot\xi + \mathbf{v}\,\xi\,\mathbf{v}\,.
\ee
An explicit calculation shows that the GKLS vector field is:

\be
\Gamma=Z_{\mathcal{K}} + Y_{\mathbf{b}}=\left(Tr\left(\mathbf{v}\,\tilde{\xi}\,\mathbf{v}\,\mathbf{e}^{k}\right) - Tr\left(\mathbf{v}^{2}\,\tilde{\xi}\,\mathbf{e}^{k}\right)\right)\,\frac{\partial}{\partial x^{k}}\,,
\ee
where we have used the notation:

\be
\xi= \frac{1}{\sqrt{n}}\mathbf{e}_{0} + \tilde{\xi}=\frac{1}{\sqrt{n}}\mathbf{e}_{0} + x^{j}\mathbf{e}_{j}\,.
\ee
Since:

\be
Tr\left(\mathbf{v}\,\xi\,\mathbf{v}\,\mathbf{e}^{k}\right)=Tr\left(\mathbf{v}\,\tilde{\xi}\,\mathbf{v}\mathbf{e}^{k}\right)+ \frac{1}{n}Tr\left(\mathbf{v}^{2}\,\mathbf{e}^{k}\right)\,,
\ee
and:

\be
Tr\left(\mathbf{v}^{2}\,\xi\,\mathbf{e}^{k}\right)=Tr\left(\mathbf{v}^{2}\,\tilde{\xi}\,\mathbf{e}^{k}\right)+ \frac{1}{n}Tr\left(\mathbf{v}^{2}\,\mathbf{e}^{k}\right)\,,
\ee
we may write the GKLS vector field $\Gamma$ as follows:

\be\label{eqn: GKLS vector field for gaussian semigroups}
\Gamma = \left(Tr\left(\mathbf{v}\,\xi\,\mathbf{v}\,\mathbf{e}^{k}\right) - Tr\left(\mathbf{v}^{2}\,\xi\,\mathbf{e}^{k}\right)\right)\,\frac{\partial}{\partial x^{k}}\,.
\ee
The fixed points of $\Gamma$ are all those $\xi_{f}$ such that:

\be
Tr\left(\mathbf{v}\,\left[\xi_{f}\,,\mathbf{v}\right]\,\mathbf{e}^{k}\right)=0\;\;\;\;\forall k=1,..n^{2}-1\,.
\ee

\begin{prop}
The function:

\be
F(\xi)=\frac{Tr\left(\xi^{2}\right)}{2}
\ee
is a LaSalle function for the GKLS vector field $\Gamma$ of equation (\ref{eqn: GKLS vector field for gaussian semigroups}).

\begin{pf}
By direct computation we have:

\be
\left.\mathcal{L}_{\Gamma}F\right|_{\rho}=\delta_{jk}x^{j}\,\left(Tr\left(\mathbf{v}\,\rho\,\mathbf{v}\,\mathbf{e}^{k}\right) - Tr\left(\mathbf{v}^{2}\,\rho\,\mathbf{e}^{k}\right)\right)=Tr\left(\mathbf{v}\,\rho\,\mathbf{v}\,\tilde{\rho}\right) - Tr\left(\mathbf{v}^{2}\,\rho\,\tilde{\rho}\right)\,,
\ee
where we have used equation (\ref{eqn: derivative of purity}) for the derivative of $F$.
Since:

\be
Tr\left(\mathbf{v}\,\rho\,\mathbf{v}\,\rho\right)=Tr\left(\mathbf{v}\,\rho\,\mathbf{v}\,\tilde{\rho}\right) + \frac{1}{n}Tr\left(\mathbf{v}\,\rho\,\mathbf{v}\right)\,,
\ee
and:

\be
Tr\left(\mathbf{v}^{2}\,\rho\,\rho\right)=Tr\left(\mathbf{v}^{2}\,\rho\,\tilde{\rho}\right) + \frac{1}{n}Tr\left(\mathbf{v}^{2}\,\rho\right)\,,
\ee
we may write:

\be
\left.\mathcal{L}_{\Gamma}F\right|_{\rho}=Tr\left(\mathbf{v}\,\rho\,\mathbf{v}\,\rho\right) - Tr\left(\mathbf{v}^{2}\,\rho^{2}\right)\,.
\ee
Note that the first and second terms in the RHS are both positive real numbers.
Now, the first term in the RHS is the inner product 

$$
\langle \rho\,\mathbf{v}|\mathbf{v}\,\rho\rangle_{\mathcal{B}(\mathcal{H})}=Tr\left((\rho\mathbf{v})^{\dagger}\,\mathbf{v}\,\rho\right)=Tr\left(\mathbf{v}\,\rho\,\mathbf{v}\,\rho\right)
$$
between $\rho\,\mathbf{v}$ and $\mathbf{v}\,\rho$.
Analogously, the second term in the RHS is the inner product $\langle \rho\,\mathbf{v}|\rho\,\mathbf{v}\rangle_{\mathcal{B}(\mathcal{H})}=|\rho\,\mathbf{v}|^{2}$ between $\rho\,\mathbf{v}$ and itself.
Being $\langle \rho\,\mathbf{v}|\mathbf{v}\,\rho\rangle_{\mathcal{B}(\mathcal{H})}$ real and positive, we may write:

\be
\langle \rho\,\mathbf{v}|\mathbf{v}\,\rho\rangle_{\mathcal{B}(\mathcal{H})}=|\rho\,\mathbf{v}|\,|\mathbf{v}\,\rho|\,\cos(\theta)\,,
\ee
where $\theta$ is the angle between $\rho\,\mathbf{v}$ and $\mathbf{v}\,\rho$.
Furthermore, since:

\be
|\mathbf{v}\,\rho|^{2}=Tr\left((\mathbf{v}\,\rho)^{\dagger}\,\mathbf{v}\rho\right)=Tr\left(\rho\,\mathbf{v}^{2}\,\rho\right)=Tr\left(\mathbf{v}^{2}\,\rho^{2}\right)=|\rho\,\mathbf{v}|^{2}\,,
\ee
we have:

\be
\left.\mathcal{L}_{\Gamma}F\right|_{\rho}=|\rho\,\mathbf{v}|\,|\mathbf{v}\,\rho|\,\cos(\theta) - |\rho\,\mathbf{v}|^{2}=|\rho\,\mathbf{v}|^{2}\left(\cos(\theta)-1\right)\leq 0\,,
\ee
and thus, theorem \ref{thm: lasalle invariance principle} applies, and the proposition is proved.
\end{pf}
\end{prop}

If we want to consider the quantum semigroup with GKLS generator:

\be\label{eqn: GKLS generator of linear combination of gaussian semigroups with hamiltonian}
\mathbf{L}(\xi)=-2\left[\left[\mathbf{H}\,,\xi\right]\right] - \mathbf{V}\odot\xi + \sum_{j=1}^{N}\,|\alpha_{j}|^{2}\mathbf{v}_{j}\,\xi\,\mathbf{v}_{j}\,,
\ee
with $\mathbf{V}=\sum_{j=1}^{N}\,|\alpha_{j}|^{2}\mathbf{v}_{j}^{2}$, we may proceed in complete analogy with what has been done for the GKLS generator of equation (\ref{eqn: GKLS generator for linear combination of quantum poisson semigroups with hamiltonian}), to obtain:

\begin{prop}\label{prop: purity is a lasalle function for linear combinations of quantum gaussian semigroups with hamiltonian}
The function:

\be
F(\xi)=\frac{Tr\left(\xi^{2}\right)}{2}
\ee
is a LaSalle function for the GKLS vector field $\Gamma$ associated with the GKLS generator of eqaution (\ref{eqn: GKLS generator of linear combination of gaussian semigroups with hamiltonian}).
\end{prop}

Essentially, the Hamiltonian term will not contribute to the Lie derivative of the function $F$, while the vector field $Y_{\mathbf{b}} + Z_{\mathcal{K}}$ will be decomposed as a positive linear combination of vector fields representing the GKLS vector fields of Gaussian semigroups, and thus, the Lie derivative of $F$ with respect to $\Gamma$ will be always negative when evaluated on the space $\stsp$ of quantum states.
This means that $F$ is a LaSalle function according to theorem \ref{thm: lasalle invariance principle}.

Similarly to what happens for quantum Poisson and quantum Gaussian semigroups, the explicit form of the accumulating set $S_{\infty}$ requires a case by case analysis.
However, the maximally mixed state will always be in $S_{\infty}$.

\subsection{Quantum random unitary semigroups}

As said before, the GKLS generator $\mathbf{L}$ for A quantum random unitary semigroups is characterized by the following form \cite{aniello_kossakowski_marmo_ventriglia-quantum_brownian_motion_on_lie_groups_and_open_quantum_systems} :

\be
\mathbf{L}(\xi)=-2[[\mathbf{H}\,,\xi]] - \mathbf{V}\odot\xi + \sum_{j=1}^{n^{2}-1}\,\alpha_{j}\mathbf{e}_{j}\,\xi\,\mathbf{e}_{j} + \beta\sum_{j=1}^{r\leq n^{2}-1}\,p_{j}\mathbf{U}_{j}\,\xi\,\mathbf{U}_{j}^{\dagger}\,,
\ee
where $\mathbf{H}\in\mathcal{B}(\mathcal{H})$ is self-adjoint, $\alpha,\beta$ are non-negative real numbers, $\{p_{j}\}_{j=1,..r}$ is a probability vector, $\{\mathbf{e}_{j}\}_{j=1,...,n^{2}-1}$ is an orthonormal set of self-adjoint operators in $\mathcal{B}(\mathcal{H})$, $\mathbf{V}=\sum_{j=1}^{n^{2}-1}\mathbf{v}^{2}$, and $U_{j}$ is unitary for all $j$.
It is clear that we may write $\mathbf{L}$ as:

\be
\mathbf{L}(\xi)= -2[[\mathbf{H}\,,\xi]] + \sum_{j=1}^{n^{2}-1}\,\alpha_{j} L_{G}^{j}(\xi) + \beta\sum_{j=1}^{r\leq n^{2}-1}\,p_{j}L_{P}^{j}(\xi)\,
\ee
where $L_{G}^{j}(\xi)$ is the GKLS generator of a quantum Gaussian semigroup, and $L_{P}^{j}(\xi)$ is the GKLS generator of a quantum Poisson semigroup.
From this decomposition, it follows that the GKLS vector field $\Gamma$ may be written as:

\be\label{eqn: GKLS vector field for quantum random unitary semigroups}
\Gamma=X_{\mathbf{a}} +  \sum_{j=1}^{n^{2}-1}\,\alpha_{j} \Gamma_{G}^{j} + \beta\sum_{j=1}^{r\leq n^{2}-1}\,p_{j}\Gamma_{P}^{j}\,
\ee
where $\Gamma_{G}^{j}$ is the GKLS vector field of the quantum Gaussian semigroup with generator $L_{G}^{j}$, and $\Gamma_{P}^{j}$ is the GKLS vector field of the quantum Poisson semigroup with generator $L_{P}^{j}$.
From this, it naturally follows the maximally mixed state is a fixed point for every such $\Gamma$ being it a fixed point for $X_{\mathbf{a}}$, $\Gamma_{G}^{j}$ and $\Gamma_{P}^{j}$.
Furthermore, we immediately have that:

\begin{prop}\label{prop: purity is a lasalle function for quantum random unitary semigroups}
The function:

\be
F(\xi)=\frac{Tr\left(\xi^{2}\right)}{2}
\ee
is a LaSalle function for the GKLS vector field $\Gamma$   of eqaution (\ref{eqn: GKLS vector field for quantum random unitary semigroups})
\begin{pf}
It follows from proposition \ref{prop: purity is a lasalle function for linear combinations of quantum poisson semigroups with hamiltonian} and proposition \ref{prop: purity is a lasalle function for linear combinations of quantum gaussian semigroups with hamiltonian}.
\end{pf}
\end{prop}

Again, the explicit form of $S_{\infty}$ requires a case by case analysis, but the maximally mixed state will always be in $S_{\infty}$.

\section{Conclusions}\label{sec: Conclusions}

We have presented here a geometric formulation of the dynamics of open quantum systems on the stratified manifold of quantum states.
Specifically, we have shown how to construct a vector field $\Gamma$ out of the algebraic equation  (\ref{eqn: K-L equation}) describing the GKLS generator $\mathbf{L}$ \cite{gorini_kossakowski_sudarshan-completely_positive_dynamical_semigroups_of_N-level_systems,lindblad-on_the_generators_of_quantum_dynamical_semigroups}.
 
The vector field $\Gamma$ is defined on the affine manifold $\mathfrak{T}_{1}$ consisting of positive, normalized linear functionals on the $C^{*}$-algebra $\appa=\mathcal{B}(\mathcal{H})$ of a finite-level quantum systems with Hilbert space $\mathcal{H}$.
By construction, $\Gamma$ is written as the linear combination of three vector fields, namely, a Hamiltonian vector field $X_{\mathbf{H}}$ associated with the Hamiltonian operator $\mathbf{H}$ in equation  (\ref{eqn: K-L equation}) by means of a Poisson tensor $\Lambda$ on $\mathfrak{T}_{1}$; a gradient-like vector field $Y_{\mathbf{V}}$ associated with the positive operator $\mathbf{V}$ in equation  (\ref{eqn: K-L equation}) by means of a symmetric bivector field $\mathcal{R}$; and a vector field $Z_{\mathcal{K}}$ associated with the completely-positive map $\mathcal{K}$ in equation  (\ref{eqn: K-L equation}).
The decomposition of $\Gamma$ is adapted to the geometry of the space $\stsp$ of quantum states when we thought of as a compact convex body in $\mathfrak{T}_{1}$.
In particular, Hamiltonian and gradient-like vector fields provide a realization of the Lie algebra $\mathfrak{sl}(\mathcal{H}\,,\mathbb{C})$ of the special linear group $SL(\mathcal{H}\,,\mathbb{C})$ on $\mathfrak{T}_{1}$, integrating to a nonlinear  action on the space $\stsp$ of quantum states.
The flow of the vector field $Z_{\mathcal{K}}$ turns out to be responsible for the change of rank of quantum states.
Interestingly, it is found that $X_{\mathbf{H}}$ is completely unrelated to the vector fields $Y_{\mathbf{V}}$ and $Z_{\mathcal{K}}$.
On the other hand, the linearity of the GKLS generator $\mathbf{L}$ requires a fine tuning between $Y_{\mathbf{V}}$ and $Z_{\mathcal{K}}$, specifically, their linear combination must be appropriately fine-tuned in order to preserve the linearity of $\mathbf{L}$.

In section \ref{sec: Quantum Poisson semigroups} the geometrical formalism presented is applied to the case of  quantum Poisson semigroups, of quantum Gaussian semigroups,
 and random unitary semigroups \cite{lindblad-on_the_generators_of_quantum_dynamical_semigroups,kossakowski-on_quantum_statistical_mechanics_of_non_hamiltonian_systems,aniello_kossakowski_marmo_ventriglia-quantum_brownian_motion_on_lie_groups_and_open_quantum_systems}.
 It is found that the vector field formalism allows to use mathematical results from the classical theory of dynamical systems in the quantum context.
By means of these tools, it is shown that every such dynamics admits an accumulating set, that is, the dynamical evolution of every quantum state $\rho$ tends to a non-equilibrium steady state $\rho_{\infty}$.
Concrete examples show that $\rho_{\infty}$ may or may not depend on the initial state $\rho$. 

We believe that the geometrical reformulation of open quantum dynamics could provide some new insights on the mathematical structure of open quantum systems, as well as the possibility of replenish the arsenal of useful mathematical tools bringing in elements from the classical theory of dynamical systems.

By making the coefficients in the module of vector fields entering the decomposition into time dependent ones, the presentation may be adapted to the description of non Markovian systems.

\section*{Acknowledgments}
We thank J. M. P\'{e}rez-Pardo for helping with the figures.\\
A. I. was partially supported by the Community of Madrid project QUITEMAD+, S2013/ICE-2801, and MINECO grant MTM2014-54692-P.\\
G. Marmo would like to acknowledge the support provided by the Banco de Santander-UCIIIM ``Chairs of Excellence'' Programme 2016-2017.

\end{document}